\documentclass[aps,prd,twocolumn,superscriptaddress,showpacs,preprintnumbers,amsmath,amssymb,tightenlines]{revtex4}

\usepackage{color}
\usepackage{graphicx} 
\usepackage{dcolumn}  
\usepackage{subfigure} 
\usepackage{hyperref} 

\graphicspath{{ps}}

\renewcommand{\arraystretch}{1.1}

\newcommand{\Bdecay}{B^0 \rightarrow \phi(K^{+}\pi^{-})^{*}}
\newcommand{\BdecayPHSP}{B^0 \rightarrow \phi K^{+}\pi^{-}}
\newcommand{\BdecayCC}{\bar{B}^0 \rightarrow \phi(K^{-}\pi^{+})^{*}}
\newcommand{\BdecayK}{B^0 \rightarrow \phi K^{*}}

\newcommand{\Swave}{(K\pi)^{*}_{0}}
\newcommand{\Pwave}{K^{*}(892)^{0}}
\newcommand{\Dwave}{K^{*}_{2}(1430)^{0}}
\newcommand{\BdecayS}{B^0 \rightarrow \phi(K\pi)^{*}_{0}}
\newcommand{\BdecayP}{B^0 \rightarrow \phi K^{*}(892)^{0}}
\newcommand{\BdecayD}{B^0 \rightarrow \phi K^{*}_{2}(1430)^{0}}
\newcommand{\Bdecayf}{B^0 \rightarrow f_0(980) K^{*}(892)^{0}}
\newcommand{\BdecayKK}{B^0 \rightarrow K^+ K^- K^{*}(892)^{0}}
\newcommand{\BdecayControl}{B^0 \rightarrow J/\psi K^{*}(892)^{0}}
\newcommand{\BdecayPhiPhi}{B^0 \rightarrow \phi \phi}
\newcommand{\BdecayPhiRho}{B^0 \rightarrow \phi \rho^0}
\newcommand{\BdecayDsK}{B^{0} \rightarrow D_s^- K^+}
\newcommand{\bbar}{B \bar{B}}

\newcommand{\mbc}{M_{\rm bc}}
\newcommand{\deltae}{\Delta E}
\newcommand{\cnb}{C_{\rm NB}'}
\newcommand{\mkk}{M_{KK}}
\newcommand{\mkpi}{M_{K\pi}}
\newcommand{\helphi}{\Phi}
\newcommand{\helthetaone}{\cos\theta_{1}}
\newcommand{\helthetaonesquared}{\cos^2\theta_{1}}
\newcommand{\helthetatwo}{\cos\theta_{2}}
\newcommand{\mphik}{M_{\phi K}}

\begin{document}

\preprint{\vbox{
					\hbox{Belle Preprint \# 2013-16}
					\hbox{KEK Preprint \# 2013-26}
}}

\title{Angular analysis of \boldmath{$\BdecayK$} decays and search for \boldmath{$CP$} violation at Belle}


\affiliation{University of the Basque Country UPV/EHU, 48080 Bilbao}
\affiliation{Beihang University, Beijing 100191}
\affiliation{Budker Institute of Nuclear Physics SB RAS and Novosibirsk State University, Novosibirsk 630090}
\affiliation{Faculty of Mathematics and Physics, Charles University, 121 16 Prague}
\affiliation{Chiba University, Chiba 263-8522}
\affiliation{Deutsches Elektronen--Synchrotron, 22607 Hamburg}
\affiliation{Department of Physics, Fu Jen Catholic University, Taipei 24205}
\affiliation{Justus-Liebig-Universit\"at Gie\ss{}en, 35392 Gie\ss{}en}
\affiliation{Hanyang University, Seoul 133-791}
\affiliation{University of Hawaii, Honolulu, Hawaii 96822}
\affiliation{High Energy Accelerator Research Organization (KEK), Tsukuba 305-0801}
\affiliation{Ikerbasque, 48011 Bilbao}
\affiliation{Indian Institute of Technology Guwahati, Assam 781039}
\affiliation{Indian Institute of Technology Madras, Chennai 600036}
\affiliation{Institute of High Energy Physics, Chinese Academy of Sciences, Beijing 100049}
\affiliation{Institute of High Energy Physics, Vienna 1050}
\affiliation{Institute for High Energy Physics, Protvino 142281}
\affiliation{INFN - Sezione di Torino, 10125 Torino}
\affiliation{Institute for Theoretical and Experimental Physics, Moscow 117218}
\affiliation{J. Stefan Institute, 1000 Ljubljana}
\affiliation{Kanagawa University, Yokohama 221-8686}
\affiliation{Institut f\"ur Experimentelle Kernphysik, Karlsruher Institut f\"ur Technologie, 76131 Karlsruhe}
\affiliation{Korea Institute of Science and Technology Information, Daejeon 305-806}
\affiliation{Korea University, Seoul 136-713}
\affiliation{Kyungpook National University, Daegu 702-701}
\affiliation{\'Ecole Polytechnique F\'ed\'erale de Lausanne (EPFL), Lausanne 1015}
\affiliation{Faculty of Mathematics and Physics, University of Ljubljana, 1000 Ljubljana}
\affiliation{Luther College, Decorah, Iowa 52101}
\affiliation{University of Maribor, 2000 Maribor}
\affiliation{Max-Planck-Institut f\"ur Physik, 80805 M\"unchen}
\affiliation{School of Physics, University of Melbourne, Victoria 3010}
\affiliation{Moscow Physical Engineering Institute, Moscow 115409}
\affiliation{Moscow Institute of Physics and Technology, Moscow Region 141700}
\affiliation{Graduate School of Science, Nagoya University, Nagoya 464-8602}
\affiliation{Kobayashi-Maskawa Institute, Nagoya University, Nagoya 464-8602}
\affiliation{National Central University, Chung-li 32054}
\affiliation{National United University, Miao Li 36003}
\affiliation{Department of Physics, National Taiwan University, Taipei 10617}
\affiliation{H. Niewodniczanski Institute of Nuclear Physics, Krakow 31-342}
\affiliation{Nippon Dental University, Niigata 951-8580}
\affiliation{Niigata University, Niigata 950-2181}
\affiliation{Osaka City University, Osaka 558-8585}
\affiliation{Pacific Northwest National Laboratory, Richland, Washington 99352}
\affiliation{Panjab University, Chandigarh 160014}
\affiliation{University of Pittsburgh, Pittsburgh, Pennsylvania 15260}
\affiliation{Research Center for Electron Photon Science, Tohoku University, Sendai 980-8578}
\affiliation{University of Science and Technology of China, Hefei 230026}
\affiliation{Seoul National University, Seoul 151-742}
\affiliation{Soongsil University, Seoul 156-743}
\affiliation{Sungkyunkwan University, Suwon 440-746}
\affiliation{School of Physics, University of Sydney, NSW 2006}
\affiliation{Tata Institute of Fundamental Research, Mumbai 400005}
\affiliation{Excellence Cluster Universe, Technische Universit\"at M\"unchen, 85748 Garching}
\affiliation{Tohoku Gakuin University, Tagajo 985-8537}
\affiliation{Tohoku University, Sendai 980-8578}
\affiliation{Department of Physics, University of Tokyo, Tokyo 113-0033}
\affiliation{Tokyo Institute of Technology, Tokyo 152-8550}
\affiliation{Tokyo Metropolitan University, Tokyo 192-0397}
\affiliation{Tokyo University of Agriculture and Technology, Tokyo 184-8588}
\affiliation{University of Torino, 10124 Torino}
\affiliation{CNP, Virginia Polytechnic Institute and State University, Blacksburg, Virginia 24061}
\affiliation{Wayne State University, Detroit, Michigan 48202}
\affiliation{Yamagata University, Yamagata 990-8560}
\affiliation{Yonsei University, Seoul 120-749}
  \author{M.~Prim}\affiliation{Institut f\"ur Experimentelle Kernphysik, Karlsruher Institut f\"ur Technologie, 76131 Karlsruhe} 
  \author{I.~Adachi}\affiliation{High Energy Accelerator Research Organization (KEK), Tsukuba 305-0801} 
  \author{H.~Aihara}\affiliation{Department of Physics, University of Tokyo, Tokyo 113-0033} 
  \author{D.~M.~Asner}\affiliation{Pacific Northwest National Laboratory, Richland, Washington 99352} 
  \author{T.~Aushev}\affiliation{Institute for Theoretical and Experimental Physics, Moscow 117218} 
  \author{A.~M.~Bakich}\affiliation{School of Physics, University of Sydney, NSW 2006} 
  \author{A.~Bala}\affiliation{Panjab University, Chandigarh 160014} 
  \author{B.~Bhuyan}\affiliation{Indian Institute of Technology Guwahati, Assam 781039} 
  \author{G.~Bonvicini}\affiliation{Wayne State University, Detroit, Michigan 48202} 
  \author{A.~Bozek}\affiliation{H. Niewodniczanski Institute of Nuclear Physics, Krakow 31-342} 
% ...
\collaboration{The Belle Collaboration}
\noaffiliation

\begin{abstract}
We report the measurements of branching fractions and $CP$ violation asymmetries in $\BdecayK$ decays obtained in an angular analysis using the full data sample of $772 \times 10^6 \bbar$ pairs collected at the $\Upsilon(4S)$ resonance with the Belle detector at the KEKB asymmetric-energy $e^+ e^-$ collider. We perform a partial wave analysis to distinguish among scalar [$\BdecayS$], vector [$\BdecayP$] and tensor [$\BdecayD$] components, and determine the corresponding branching fractions to be $\mathcal{B}[\BdecayS] = (4.3 \pm 0.4 \pm 0.4) \times 10^{-6}$, $\mathcal{B}[\BdecayP] = (10.4 \pm 0.5 \pm 0.6) \times 10^{-6}$ and $\mathcal{B}[\BdecayD] = (5.5 ^{+0.9}_{-0.7} \pm 1.0) \times 10^{-6}$. We also measure the longitudinal polarization fraction $f_L$ in $\BdecayP$ and $\BdecayD$ decays to be $0.499 \pm 0.030 \pm 0.018$ and $0.918 ^{+0.029}_{-0.060} \pm 0.012$, respectively. The first quoted uncertainties are statistical and the second are systematic. In total, we measure 26 parameters related to branching fractions, polarization and $CP$ violation in the $\BdecayK$ system. No evidence for $CP$ violation is found.
\end{abstract}


\maketitle

{\renewcommand{\thefootnote}{\fnsymbol{footnote}}}
\setcounter{footnote}{0}

\section{Introduction}
In the standard model (SM) of electroweak interactions, the effect of $CP$ violation is explained by a single irreducible phase in the $3\times 3$ Cabibbo-Kobayashi-Maskawa (CKM) quark mixing matrix~\cite{C,KM}. So far, analyses~\cite{HFAG} searching for $CP$ violation have shown no significant deviation with respect to the SM predictions.

\begin{figure}[ht]
 \includegraphics[width=0.9\linewidth]{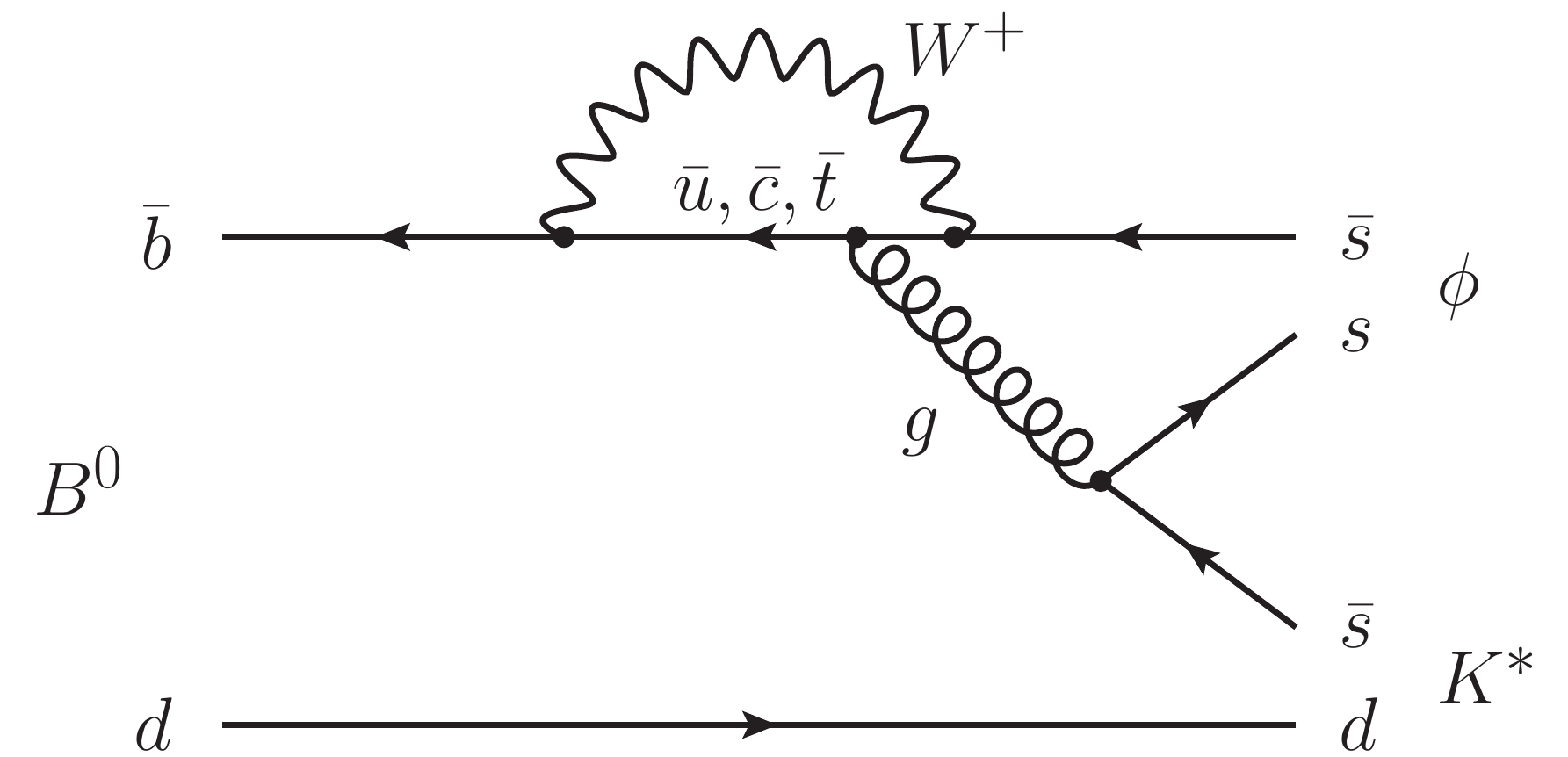}
 \caption{Penguin diagram of the decay $\BdecayK$.}
\label{fig:feynman_diagram}
\end{figure}

The CKM mechanism alone is not sufficient to explain the observed matter-antimatter asymmetry in the universe, and thus new sources of $CP$ violation are necessary. Decays dominated by $b \to s$ penguin (loop) transitions in the SM, such as $\BdecayK$, as shown in Fig.~\ref{fig:feynman_diagram}, are sensitive to such new contributions. New particles could appear in virtual loops, resulting in significant deviations from the SM expectations of negligible direct $CP$ violation. Previous studies by Belle~\cite{Belle_phiK} and BaBar~\cite{BaBar_phiK} in $\BdecayP$ did not find any evidence for $CP$ violation. On the other hand, the longitudinal polarization fractions $f_L = 0.45 \pm 0.05 \pm 0.02$ (Belle) and $f_L = 0.494 \pm 0.034 \pm 0.013$ (BaBar) in this decay were found to deviate from a naive expectation based on the factorization approach~\cite{Polarization}, which predicts a longitudinal polarization fraction close to unity. In contrast, BaBar measured the longitudinal polarization fraction in $\BdecayD$ to be $f_L = 0.901 ^{+0.046}_{-0.058} \pm 0.037$~\cite{BaBar_phiK}, consistent with the factorization prediction.

In this paper, we present an improved analysis of the $\BdecayK$~\cite{CC} system using the full Belle data sample collected at the $\Upsilon(4S)$ resonance. We perform a partial wave analysis to distinguish among the different $K^*$ states. Overall, 26 parameters related to branching fractions, polarization, interference effects and $CP$ violation are measured.

The measurement of polarization in flavor specific $\BdecayK$ decays can be used further to distinguish between $CP$-even and -odd fractions in the decay  $B^{0}/\bar{B}^{0} \to \phi K_S^0 \pi^0$. This decay channel can also be used for a time-dependent measurement of the angle $\phi_1 = \arg \left(- V_{cd}V^*_{cb}/V_{td}V^*_{tb} \right)$~\cite{alpha} of the CKM unitarity triangle in $b \to (s\bar{s})s$ transitions.

\section{Analysis strategy}

We perform a partial wave analysis of the $\BdecayK$ system with $\phi \to K^+K^-$ and $K^* \to K^+\pi^-$. We use the $K^{*}$ notation to indicate all possible contributions from scalar (S-wave, spin $J=0$), vector (P-wave, $J=1$) and tensor (D-wave, $J=2$) components from $\Swave$, $\Pwave$ and $\Dwave$, respectively. We assume no further resonant contributions. The analysis region is limited to a $K^+\pi^-$ invariant mass below $1.55$~GeV, as the LASS model~\cite{LASS}, used to parametrize the S-wave contribution, is not valid above this value. Furthermore, no significant contribution from $K^*$ states beyond $1.55$~GeV is observed~\cite{BaBar_highmass_states}. We use mass and angular distributions to distinguish among the three contributing channels $\BdecayS$, $\BdecayP$, and $\BdecayD$, and to determine the polarization in vector--vector and vector--tensor decays, as well as a number of parameters related to $CP$ violation. We also determine the branching fraction for each of the three channels.

We first explain the parametrization of the angular distribution, which is followed by a description of the $K^{+}\pi^{-}$ invariant-mass distribution. Finally, we derive the combined model of mass and angular distributions of partial waves used for the parameter extraction in a maximum likelihood fit.

\subsection{Angular distribution}

\begin{figure}[ht]
 \includegraphics[width=0.9\linewidth]{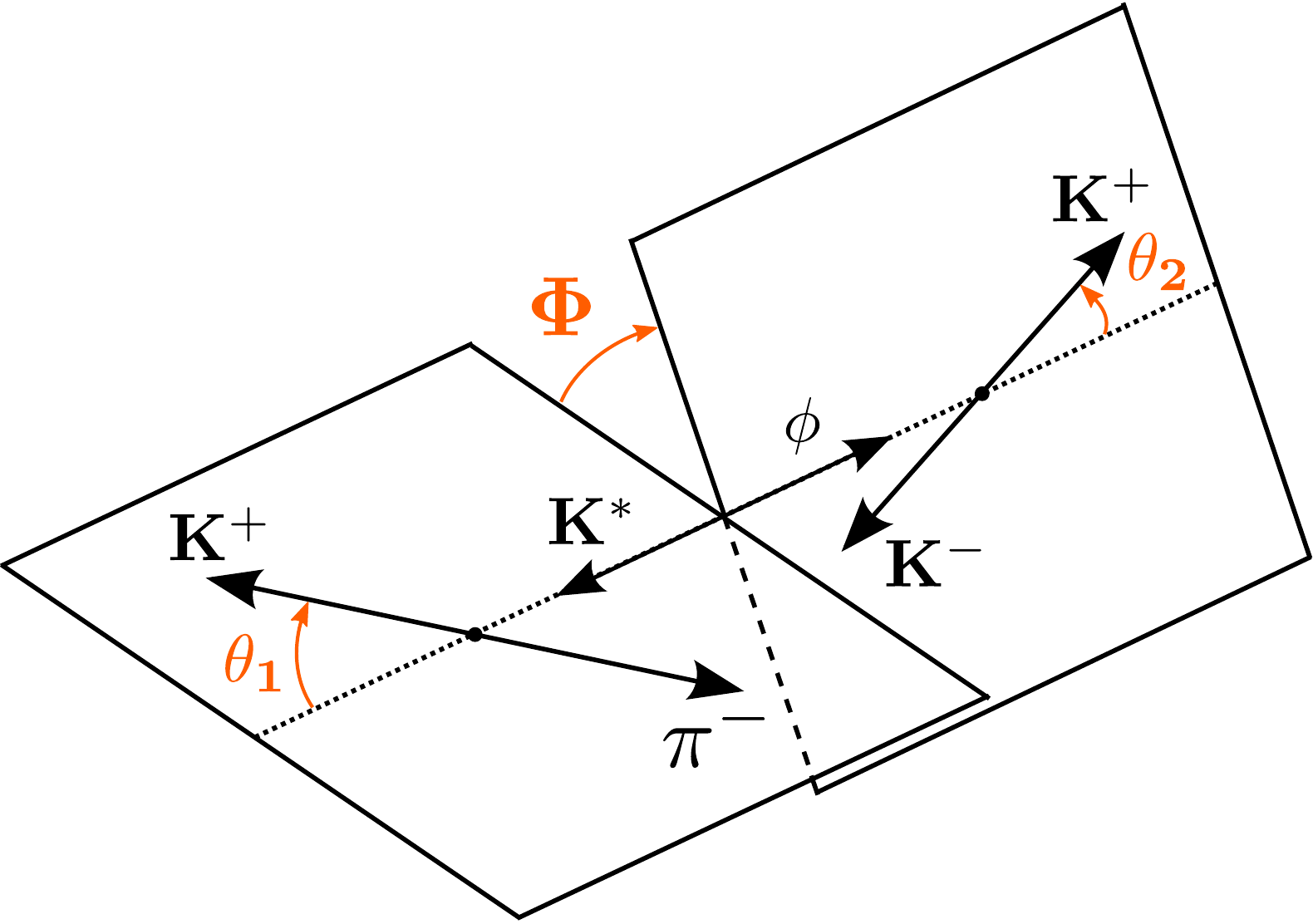}
 \caption{Definition of the three helicity angles given in the rest frame of the parent particles for the $\BdecayK$ decay.}
\label{fig:helicity_angles}
\end{figure}

The angular distribution in the $\BdecayK$ system with $\phi \to K^{+} K^{-}$ and $K^* \to K^+ \pi^-$ is described by the three helicity angles $\theta_1$, $\theta_2$, and $\Phi$, which are defined in the rest frame of the parent particles as illustrated in Fig.~\ref{fig:helicity_angles}.

In general, due to the angular momentum conservation, the partial decay width for a two-body decay of a pseudoscalar $B$ meson into particles with spins $J_1$ and $J_2$ is given by
\begin{equation}
\frac{d^3 \Gamma}{d\cos \theta_1 d\cos \theta_2 d\Phi} \propto \left| \sum_\lambda A_\lambda Y_{J_1}^{\lambda}\left(\theta_1,\Phi\right) Y_{J_2}^{-\lambda}\left(-\theta_2,0\right)\right|^2,
\label{eqn:partial_angular_decay_width}
\end{equation}
where $Y_l^m$ are the spherical harmonics, the sum is over the helicity states $\lambda$, and $A_\lambda$ is the complex weight of the corresponding helicity amplitude. The parameter $\lambda$ takes all discrete values between $-j$ and $+j$, with $j$ being the smaller of the two daughter particle spins $J_1$ and $J_2$. As the $\phi$ is a vector meson, $J_2 = 1$ in this analysis, whereas $J_1 = 0$ for $\Swave$, $J_1 = 1$ for $\Pwave$, and $J_1 = 2$ for $\Dwave$. The partial decay width of each partial wave with spin $J \equiv J_1$ is therefore
\begin{equation}
\frac{d^3 \Gamma}{d\cos \theta_1 d\cos \theta_2 d\Phi} \propto \left| \sum_\lambda A_{J\lambda} Y_{J}^{\lambda}\left(\theta_1,\Phi\right) Y_{1}^{-\lambda}\left(-\theta_2,0\right)\right|^2,
\label{eqn:partial_angular_decay_width_per_spin}
\end{equation}
with $A_{J\lambda}$ being the complex weight of the corresponding helicity amplitude of the partial wave with spin $J$.

The helicity basis is not a basis of $CP$ eigenstates. Polarization measurements are commonly performed in the transversity basis of $CP$ eigenstates with the transformation $A_{J\pm1} = (A_{J\parallel} \pm A_{J\perp})/\sqrt{2}$ for two of the amplitudes. In this basis, the longitudinal polarization $A_{J0}$ and the parallel polarization $A_{J\parallel}$ are even under $CP$ transformation while the perpendicular component $A_{J\perp}$ is $CP$-odd. Throughout this article, we use $A$ for $B^0$ and $\bar{A}$ for $\bar{B}^0$ related complex weights of the helicity and transversity amplitudes. Furthermore, depending on the context, we use either of the two bases with $\lambda = -1,0,+1$ or $\lambda=0,\parallel,\perp$. Where necessary, we explicitly state the basis used. We use polar coordinates to define the complex weights $A_{J\lambda} = a_{J\lambda} \exp (i\varphi_{J\lambda})$ and apply the same implicit definition of the basis; e.\,g. $a_{2\perp}$ would be the magnitude of the perpendicular D-wave component in the transversity basis.

\subsection{Mass distribution}

To distinguish among different partial waves, we study their $K^{+}\pi^{-}$ invariant-mass spectrum $\mkpi$. To parametrize the lineshape of the P- and D-wave components as a function of the invariant mass $m$, we use a relativistic spin-dependent Breit--Wigner (BW) amplitude $R_J$~\cite{PDG}:
\begin{equation}
R_J(m) = \frac{m_J \Gamma_J(m)}{(m_J^2 -m^2) - im_J \Gamma_J(m)} = \sin \delta_J e^{i\delta_J},
\end{equation}
where we use the convention
\begin{equation}
\cot \delta_J = \frac{m_J^2 -m^2}{m_J\Gamma_J(m)}.
\end{equation}
For spin $J=1$ and $J=2$, the mass-dependent widths are given by
\begin{align}
\Gamma_1(m) &= \Gamma_1 \frac{m_1}{m} \frac{1+r^2 q_1^2}{1+r^2 q^2} \left( \frac{q}{q_1} \right)^3, \\
\Gamma_2(m) &= \Gamma_2 \frac{m_2}{m} \frac{9+3r^2 q_2^2 + r^4 q_2^4}{9+3r^2 q^2 + r^4 q^4} \left( \frac{q}{q_2} \right)^5,
\end{align}
where $\Gamma_J$ is the resonance width, $m_J$ the resonance mass, $q$ the momentum of a daughter particle in the rest frame of the resonance, $q_J$ this momentum evaluated at $m = m_J$, and $r$ the interaction radius. This parametrization of the mass-dependent width uses the Blatt--Weisskopf penetration factors~\cite{PDG}.

The S-wave component is parametrized using $K\pi$ scattering results from the LASS experiment~\cite{LASS}. It was found by LASS that the scattering is elastic up to about $1.5-1.6$~GeV and thus can be parametrized as
\begin{equation}
R_0(m) = \sin \delta_0 e^{i\delta_0},
\end{equation}
where
\begin{equation}
\delta_0 = \Delta R + \Delta B,
\end{equation}
$\Delta R$ representing a resonant contribution from $K_0^*(1430)^0$ while $\Delta B$ denoting a non-resonant contribution. The resonant part is defined as
\begin{equation}
\cot \Delta R = \frac{m_0^2 -m^2}{m_0\Gamma_0(m)},
\end{equation}
where $m_0$ and $\Gamma_0$ are the resonance mass and width, and $\Gamma_0(m)$ is given by
\begin{equation}
\Gamma_0(m) = \Gamma_0 \frac{m_0}{m} \left( \frac{q}{q_0} \right).
\end{equation}
The non-resonant part is defined as
\begin{equation}
\cot \Delta B = \frac{1}{aq} + \frac{bq}{2},
\end{equation}
where $a$ is the scattering length and $b$ is the effective range.

The amplitude $M_J(m)$ is obtained by multiplying the lineshape with the two-body phase space factor
\begin{equation}
M_J(m) = \frac{m}{q} R_J(m).
\label{eqn:mass_distribution_per_spin}
\end{equation}

The resonance parameters used in the analysis are given in Table~\ref{tab:resonance_parameters_Kpi}.

\begin{table}[ht]
\caption{Resonance parameters for S-, P-, and D-wave components. The parameters $m_J$ and $\Gamma_J$ for P- and D-wave are taken from Ref.~\cite{PDG}, and interaction radii and S-wave parameters are taken from Ref.~\cite{BaBar_phiK}, which includes updated values with respect to Ref.~\cite{LASS}.}
\label{tab:resonance_parameters_Kpi}
\begin{tabular}{cccc} \hline \hline
 & $\Swave$ & $\Pwave$ & $\Dwave$ \\
Parameter & $J=0$ & $J=1$ & $J=2$\\ \hline
$m_J$ (MeV) & $1435 \pm 5 \pm 5$ & $895.94 \pm 0.22$ & $1432.4 \pm 1.3$\\
$\Gamma_J$ (MeV) & $279 \pm 6 \pm 21$ & $48.7 \pm 0.8$ & $109 \pm 5$\\
$r$ (GeV$^{-1}$) & $\cdots$ & $3.4 \pm 0.7$ & $2.7 \pm 1.3$\\
$a$ (GeV$^{-1}$) & $1.95 \pm 0.09 \pm 0.06$ & $\cdots$ & $\cdots$\\
$b$ (GeV$^{-1}$) & $1.76 \pm 0.36 \pm 0.67$ & $\cdots$ & $\cdots$\\
\hline \hline 
\end{tabular}
\end{table}

\subsection{Mass-angular distribution}
We combine the mass distribution with the angular distribution to obtain the partial decay width
\begin{equation}
\frac{d^4 \Gamma}{d\cos \theta_1 d\cos \theta_2 d\Phi d\mkpi} \propto |\mathcal{M}|^2 \times F_{\mphik}\left(\mkpi\right),
\end{equation}
where $F_{\mphik}\left(\mkpi\right)$ is a phase space factor that takes into account the three-body kinematics in $B^{0} \to \phi K^+ \pi^-$. As we expect no resonant charmless structure in the $\phi K^+$ invariant-mass distribution, we assume a constant amplitude that can be computed for each value of $\mkpi$ following the section on kinematics in Ref.~\cite{PDG} as
\begin{equation}
F(m) = 2 m \left[ m^2_{\rm max}(m) - m^2_{\rm min}(m) \right],
\end{equation}
with $m^2_{\rm max}$ ($m^2_{\rm min}$) being the maximum (minimum) value of the Dalitz plot range of the $\phi K^+$ invariant mass $\mphik$ at a given $\mkpi$ value. 

\renewcommand{\arraystretch}{1.2}
\begin{table*}[ht]
\caption{Definitions of the 26 real parameters that are measured in the $\BdecayK$ system. Three partial waves with spin $J=0,1,2$ are considered in the $K^{+}\pi^{-}$ spectrum. The amplitude weights $A_{J\lambda}$ are defined in the text. The extra $\pi$ in the definition of $\phi_{\perp J}$ and $\Delta \phi_{\perp J}$ accounts for the sign flip of $A_{J\perp} = -\bar{A}_{J\perp}$ under $CP$ transformation.}
\label{tab:physics_parameter_definition}
\begin{tabular}{ccccc} \hline \hline
& & $\phi \Swave$ & $\phi \Pwave$ & $\phi \Dwave$ \\
Parameter &  Definition & $J=0$ & $J=1$ & $J=2$\\ \hline
$\mathcal{B}_J$ & $\frac{1}{2} (\bar{\Gamma}_J + \Gamma_J) / \Gamma_{\text{total}}$ & $\mathcal{B}_0$ & $\mathcal{B}_1$ & $\mathcal{B}_2$\\
$f_{LJ}$ & $\frac{1}{2} (\vert\bar{A}_{J0}\vert^2 / \sum\vert\bar{A}_{J\lambda}\vert^2 + \vert A_{J0}\vert^2 / \sum\vert A_{J\lambda}\vert^2)$ & $\cdots$ & $f_{L1}$ & $f_{L2}$\\
$f_{\perp J}$ & $\frac{1}{2}( \vert\bar{A}_{J\perp}\vert^2 / \sum\vert\bar{A}_{J\lambda}\vert^2 + \vert A_{J\perp}\vert^2 / \sum\vert A_{J\lambda}\vert^2)$ & $\cdots$ & $f_{\perp 1}$ & $f_{\perp 2}$\\
$\phi_{\parallel J}$ & $\frac{1}{2}(\arg( \bar{A}_{J\parallel} / \bar{A}_{J0}) + \arg(A_{J\parallel} / A_{J0}))$ & $\cdots$  & $\phi_{\parallel 1}$ & $\phi_{\parallel 2}$\\
$\phi_{\perp J}$ & $\frac{1}{2}(\arg(\bar{A}_{J\perp} / \bar{A}_{J0}) + \arg(A_{J\perp} / A_{J0}) - \pi)$ & $\cdots$ & $\phi_{\perp 1}$ & $\phi_{\perp 2}$\\
$\delta_{0J}$ & $\frac{1}{2}(\arg(\bar{A}_{00} / \bar{A}_{J0}) + \arg(A_{00}/A_{J0}))$ & $\cdots$ & $\delta_{01}$ & $\delta_{02}$\\
$\mathcal{A}_{CPJ}$ & $(\bar{\Gamma}_J - \Gamma_J) / (\bar{\Gamma}_J + \Gamma_J)$ & $\mathcal{A}_{CP0}$ & $\mathcal{A}_{CP1}$ & $\mathcal{A}_{CP2}$\\[3pt]
$\mathcal{A}_{CPJ}^0$ & $\frac{\vert\bar{A}_{J0}\vert^2 / \sum\vert\bar{A}_{J\lambda}\vert^2 - \vert A_{J0}\vert^2 / \sum\vert A_{J\lambda}\vert^2}{\vert\bar{A}_{J0}\vert^2 / \sum\vert\bar{A}_{J\lambda}\vert^2 + \vert A_{J0}\vert^2 / \sum\vert A_{J\lambda}\vert^2}$ & $\cdots$ & $\mathcal{A}_{CP1}^0$ & $\mathcal{A}_{CP2}^0$\\[3pt]
$\mathcal{A}_{CPJ}^\perp$ & $\frac{\vert\bar{A}_{J\perp}\vert^2 / \sum\vert\bar{A}_{J\lambda}\vert^2 - \vert A_{J\perp}\vert^2 / \sum\vert A_{J\lambda}\vert^2}{\vert\bar{A}_{J\perp}\vert^2 / \sum\vert\bar{A}_{J\lambda}\vert^2 + \vert A_{J\perp}\vert^2 / \sum\vert A_{J\lambda}\vert^2}$ & $\cdots$ & $\mathcal{A}_{CP1}^\perp$ & $\mathcal{A}_{CP2}^\perp$\\[3pt]
$\Delta\phi_{\parallel J}$ & $\frac{1}{2}(\arg(\bar{A}_{J\parallel} /\bar{A}_{J0}) - \arg(A_{J\parallel} /A_{J0}))$ & $\cdots$ & $\Delta\phi_{\parallel 1}$ & $\Delta\phi_{\parallel 2}$\\
$\Delta\phi_{\perp J}$ & $\frac{1}{2}(\arg(\bar{A}_{J\perp} /\bar{A}_{J0}) - \arg(A_{J\perp} /A_{J0}) - \pi)$ & $\cdots$ & $\Delta\phi_{\perp 1}$ & $\Delta\phi_{\perp 2}$\\
$\Delta\delta_{0J}$ & $\frac{1}{2}(\arg(\bar{A}_{00} /\bar{A}_{J0}) - \arg(A_{00} /A_{J0}))$ & $\cdots$ & $\Delta\delta_{01}$ & $\Delta\delta_{02}$\\
\hline \hline 
\end{tabular}
\end{table*}
\renewcommand{\arraystretch}{1.1}

The matrix element squared $\vert\mathcal{M}\vert^2$ is given by the coherent sum of the corresponding S-, P-, and D-wave amplitudes $A_J$ as
\begin{equation}
\begin{split}
\vert\mathcal{M}\vert^2 = \vert &\mathcal{A}_0\left(\mkpi,\helthetaone,\helthetatwo,\helphi\right)\\
+ &\mathcal{A}_1\left(\mkpi,\helthetaone,\helthetatwo,\helphi\right)\\
+ &\mathcal{A}_2\left(\mkpi,\helthetaone,\helthetatwo,\helphi\right) \vert^2,
\end{split}
\end{equation}
where we have omitted the explicit dependence of $\mathcal{M}$ on $\left(\mkpi,\helthetaone,\helthetatwo,\helphi\right)$ for readability. Each partial wave for a given spin $J$ is parametrized as the product of the angular distribution from Eq.~(\ref{eqn:partial_angular_decay_width_per_spin}) and the mass distribution from Eq.~(\ref{eqn:mass_distribution_per_spin}). For the S-, P-, and D-wave, we obtain
\begin{multline}
\label{eqn:Swave}
\mathcal{A}_0\left(\mkpi,\helthetaone,\helthetatwo,\helphi\right) \\ = A_{00} Y_0^0(\theta_1,\helphi) Y_1^0(-\theta_2,0) \times M_0(\mkpi),
\end{multline}
\begin{multline}
\label{eqn:Pwave}
\mathcal{A}_1\left(\mkpi,\helthetaone,\helthetatwo,\helphi\right) \\ = \sum_{\lambda = 0,\pm 1}A_{1\lambda} Y_1^{\lambda}(\theta_1,\helphi) Y_1^{-\lambda}(-\theta_2,0) \times M_1(\mkpi),
\end{multline}
and
\begin{multline}
\label{eqn:Dwave}
\mathcal{A}_2\left(\mkpi,\helthetaone,\helthetatwo,\helphi\right) \\ = \sum_{\lambda = 0,\pm 1}A_{2\lambda} Y_2^{\lambda}(\theta_1,\helphi) Y_1^{-\lambda}(-\theta_2,0) \times M_2(\mkpi),
\end{multline}
respectively.

Overall, the seven complex helicity amplitudes contributing to these formulas can be parametrized by 14 real parameters (28 if $B^{0}$ and $\bar{B}^{0}$ are measured independently). 

We define the normalized partial decay width as
\begin{multline}
\label{eqn:width_signal}
\frac{d^4 \Gamma}{d\cos \theta_1 d\cos \theta_2 d\Phi d\mkpi} = F_{\mphik}\left(\mkpi\right) \\
 \times \frac{(1+Q) \times \vert\mathcal{M}^+\vert^2 + (1-Q) \times \vert\mathcal{M}^-\vert^2}{2\mathcal{N}},
\end{multline}
where $\mathcal{M}^+$ [$\mathcal{M}^-$] is the matrix element for $\Bdecay$ [$\BdecayCC$], $Q$ is $\pm 1$ depending on the charge of the primary charged kaon from the $B$ meson and $\mathcal{N}$ is the overall normalization given by
\begin{equation}
\label{eqn:normalization}
\begin{split}
\mathcal{N} =\ &\frac{1}{2} \int \vert \mathcal{M}^+ \vert^2 \times F_{\mphik}\left(\mkpi\right) d\cos \theta_1 d\cos \theta_2 d\Phi d\mkpi \\ + &\frac{1}{2} \int \vert \mathcal{M}^- \vert^2 \times F_{\mphik}\left(\mkpi\right) d\cos \theta_1 d\cos \theta_2 d\Phi d\mkpi.
\end{split}
\end{equation}

By averaging the normalization over $B^{0}$ and $\bar{B}^{0}$, we can perform a simultaneous fit with a single reference amplitude of fixed magnitude, which defines the relative strengths of the amplitudes. If both final states are normalized independently, each with its own reference amplitude, and $CP$ violation is observed, the interpretation of whether $CP$ violation is in the reference amplitudes or all other amplitudes would be ambiguous.

Using these notations, we define the final set of parameters used in the analysis. For the matrix element $\mathcal{M}^+$, we define the weights as $A_{J\lambda} = a_{J\lambda}^+ \exp (i\varphi_{J\lambda}^+)$ and, for $\mathcal{M}^-$, as $\bar{A}_{J\lambda} = a_{J\lambda}^- \exp (i\varphi_{J\lambda}^-)$. With $a_{J\lambda}^\pm$ defined as
\begin{equation}
\label{eqn:fit_parameters}
a_{J\lambda}^\pm = a_{J\lambda}  (1 \pm \Delta a_{J\lambda})
\end{equation}
and $\varphi_{J\lambda}^\pm$ given by
\begin{equation}
\label{eqn:fit_parameters_end}
\varphi_{J\lambda}^\pm = \varphi_{J\lambda} \pm \Delta \varphi_{J\lambda},
\end{equation}
where we use one $CP$-conserving and one $CP$-violating parameter per magnitude and phase. For $J=0$ only $\lambda=0$ is possible, whereas, for $J=1$ and $J=2$, three values $\lambda=0,\parallel$ and $\perp$ are allowed.

We choose $\varphi_{00} = 0$ as our reference phase, as the system is invariant under a global phase transformation. This effectively reduces the 28 parameters by one. Of the remaining 27 parameters, 26 can be measured in the $\BdecayK$ system with $K^* \to K^+ \pi^-$. These 26 parameters can be used to define a more common set of parameters shown in Table~\ref{tab:physics_parameter_definition}, which are used in the review of polarization in $B$ decays in Ref.~\cite{PDG}. For each partial wave $J$, we define parameters such as the longitudinal (perpendicular) polarization fractions $f_{LJ}$ ($f_{\perp J}$), the relative phase of the parallel (perpendicular) amplitude $\phi_{\parallel J}$ ($\phi_{\perp J}$) to the longitudinal amplitude, and strong phase difference between the partial waves $\delta_{0J}$ and a number of parameters related to $CP$ violation. The 27\textsuperscript{th} parameter, $\Delta \varphi_{00} = \Delta \phi_{00} = \frac{1}{2}\arg(A_{00} /\bar{A}_{00})$, could only be measured in a time-dependent analysis of $CP$ violation in $B^{0}/\bar{B}^{0} \to \phi K_S^0 \pi^0$ decays that is beyond the scope of this analysis, so we fix $\Delta \varphi_{00} = 0$. Furthermore, we fix $a_{10}$ as it has the largest relative magnitude among all amplitudes and choose it as our reference amplitude. Fixing $a_{10}$ does not decrease the number of free parameters as the absolute magnitude, defined by the signal yield, remains a free parameter in the fit. Overall, we are left with 26 real parameters to be determined.

In the previous analysis~\cite{Belle_phiK}, a twofold phase ambiguity was observed in the decay of $\BdecayP$; this is a fourfold ambiguity if $B^0$ and $\bar{B}^0$ are measured independently, as the sets $(\phi_{\parallel J}, \phi_{\perp J}, \Delta \phi_{\parallel J}, \Delta \phi_{\perp J})$ and $(2\pi - \phi_{\parallel J}, \pi - \phi_{\perp J}, -\Delta \phi_{\parallel J}, -\Delta \phi_{\perp J})$ solve all angular equations.  Even the interference terms in $\vert \mathcal{M} \vert^2$ are invariant under such transformation if we flip the sign of the strong phase $\delta_{0J}$. However, the mass dependence of $\delta_{0J}$ is unique: it either increases or decreases with increasing $K^+\pi^-$ invariant mass. We solve this ambiguity for $B^0$ and $\bar{B}^0$ using Wigner's causality principle~\cite{Wigner}, which states that the phase of a resonance increases with increasing invariant mass. 

From the measured weights, we can also calculate the triple-product correlations in $\BdecayP$, given in our previous measurement. The $T$-odd quantities
\begin{equation}
A_T^0 = \operatorname{Im}( A_{1\perp} A_{10}^* ) \text{\quad and \quad} A_T^\parallel = \operatorname{Im} ( A_{1\perp} A_{1\parallel}^* )
\end{equation}
from Ref.~\cite{TripleProduct,TripleProduct_note} and the corresponding asymmetries $\mathcal{A}_{T}^{0/\parallel}$ between $B^0$ and $\bar{B}^0$ are sensitive to $T$-odd $CP$ violation.

\section{Event reconstruction}

\subsection{Data sample and detector}

We use the full Belle data sample, consisting of an integrated luminosity of $711\,{\rm fb}^{-1}$ containing $(772 \pm 11) \times 10^6 \bbar$ pairs collected at the $\Upsilon(4S)$ resonance at the KEKB asymmetric-energy $e^+e^-$ (3.5 on 8~GeV) collider~\cite{KEKB}. An additional data sample of $79\,{\rm fb}^{-1}$ integrated luminosity collected 60~MeV below the $\Upsilon(4S)$ resonance, referred to as the off-resonance data, is utilized for background studies.

The Belle detector~\cite{Belle} is a large-solid-angle magnetic
spectrometer that consists of a silicon vertex detector,
a 50-layer central drift chamber (CDC), an array of
aerogel threshold Cherenkov counters (ACC),
a barrel-like arrangement of time-of-flight
scintillation counters (TOF), and an electromagnetic calorimeter
composed of CsI(Tl) crystals located inside 
a superconducting solenoid coil that provides a 1.5~T
magnetic field.  An iron flux-return located outside of
the coil is instrumented to detect $K_L^0$ mesons and to identify
muons.
Two inner detector configurations were used. A 2.0 cm beampipe
and a 3-layer silicon vertex detector were used for the first sample
of $152 \times 10^6 B\bar{B}$ pairs, while a 1.5 cm beampipe, a 4-layer
silicon detector and a small-cell inner drift chamber were used to record  
the remaining $620 \times 10^6 B\bar{B}$ pairs~\cite{svd2}.

\subsection{Event reconstruction and selection}

We reconstruct $B^0$ candidates in the decay mode $B^{0} \to \phi K^+ \pi^-$ with $\phi \to  K^+ K^-$. The charged tracks are required to have a transverse (longitudinal) distance of closest approach to the interaction point (IP) of less than $0.1$ $(4.0)$~cm. For particle identification (PID) of track candidates, specific energy loss measured in the CDC and information from the ACC and the TOF are combined using a likelihood-ratio approach. The selection requirement on the combined PID quantity has a kaon (pion) identification efficiency of $95\%$ $(98\%)$ with an associated pion (kaon) misidentification rate of $26\%$ $(9\%)$ for the track candidates not used as primary kaon from the $B$ meson. For a primary kaon from the $B$ meson candidate, the kaon identification efficiency is $90\%$ with an associated pion misidentification rate of $28\%$. The $K^+ K^-$ invariant mass for $\phi$ candidates is required to be $\mkk < 1.05$~GeV. The $K^+ \pi^-$ invariant mass must satisfy the criterion $0.7\text{~GeV} < \mkpi < 1.55\text{~GeV}$. 

The selection of $B^0$ candidates is based on the beam-energy-constrained mass $\mbc = \sqrt{(E_{\rm beam}^*)^2 - (p_B^*)^2}$ and the energy difference $\deltae = E_B^* - E_{\rm beam}^*$, where $E_{\rm beam}^*$ is the beam energy, and $p_B^*$ and $E_B^*$ are the momentum and energy of the $B^0$ candidates in the center-of-mass (CM) frame, respectively. Candidates with $5.24 \text{~GeV} < \mbc < 5.29 \text{~GeV}$ and $-150 \text{~MeV} < \deltae < 150 \text{~MeV}$ are retained for further analysis. The range $5.24 \text{~GeV} < \mbc < 5.26 \text{~GeV}$ is used as the sideband, whereas $5.26 \text{~GeV} < \mbc < 5.29 \text{~GeV}$ is used as the nominal fit region.

In 17\% of all signal events, more than one $B^0$ candidate passes the above selection; we select the candidate with the smallest $\chi^2$ for the hypothesis that all tracks form a common vertex within the IP region. This requirement selects the correct candidate with a probability of 64\% according to Monte Carlo (MC) simulations.

The dominant background arises from $e^+e^- \to q\bar{q} \ (q \in \left\lbrace u,d,s,c \right\rbrace)$ continuum events, which are suppressed using a neural network (NN) implemented with the NeuroBayes package~\cite{NeuroBayes}. In the NN, we combine $\cos \theta_B$, the polar angle of the $B^0$ candidate with respect to the beam direction in the CM frame, a likelihood constructed from 16 modified Fox--Wolfram moments~\cite{SFW} and $\cos \theta_T$, the polar angle between the thrust axis of the $B^0$ candidate and the remaining tracks in the event. The NN assigns each candidate a value, $C_{\rm NB}$, in the interval $[-1,+1]$ with $-1$ $(+1)$ being background (signal)-like. We require $C_{\rm NB}>0$ to reject 86\% of the background while retaining 83\% of the signal. Hereinafter, we refer to the continuum background, together with a 2\% contribution from random combinations of tracks from $\bbar$ events, as the combinatorial background.

The remaining background contribution arises from $\bbar$ events and is due either to signal events in which we select a candidate with at least one track originating from the other $B$ [referred to as self-crossfeed (SCF)], or peaking background from $B$ decays. The SCF events are mainly due to partially reconstructed $B^0$ candidates, with a $\pi^-$ track from the other $B$ meson. Often, the pion momentum is low compared to the kaon momentum so that the direction of the $K^+\pi^-$ system is dominated by the $K^+$ momentum. These combinations tend to peak in the region of high $\cos \theta_1$ values. The peaking background originates from either $\BdecayDsK$ with $D_s^- \to \phi \pi^-$, which peaks sharply near $0.8$ in the $\cos \theta_1$ distribution, or from $\Bdecayf$ events. We require $\cos \theta_1 < 0.75$ to reject the peaking $\BdecayDsK$ events completely as well as a majority of the SCF events. With respect to signal, about 5\% of the events are due to SCF that will be discussed further in Sec.~\ref{sec:systematics}.

The reconstruction and selection procedures are established using MC events generated with the EvtGen program~\cite{EvtGen} and a full detector simulation based on GEANT3~\cite{GEANT3}. The PHOTOS package~\cite{PHOTOS} is used to take into account final state radiation. The MC statistics for CKM-favored $b \to c$ transitions and $q\bar{q}$ decays correspond to four times the data statistics. In addition, we use an MC sample of rare $b \to s$ decays with 50 times the statistics of the data sample. We further use a very large sample of $\BdecayPHSP$ three-body phase space decays for our studies and several samples with different polarizations for cross-checks.

\subsection{Efficiency}
\begin{figure}[ht]
\centering
 \includegraphics[width=0.49\linewidth]{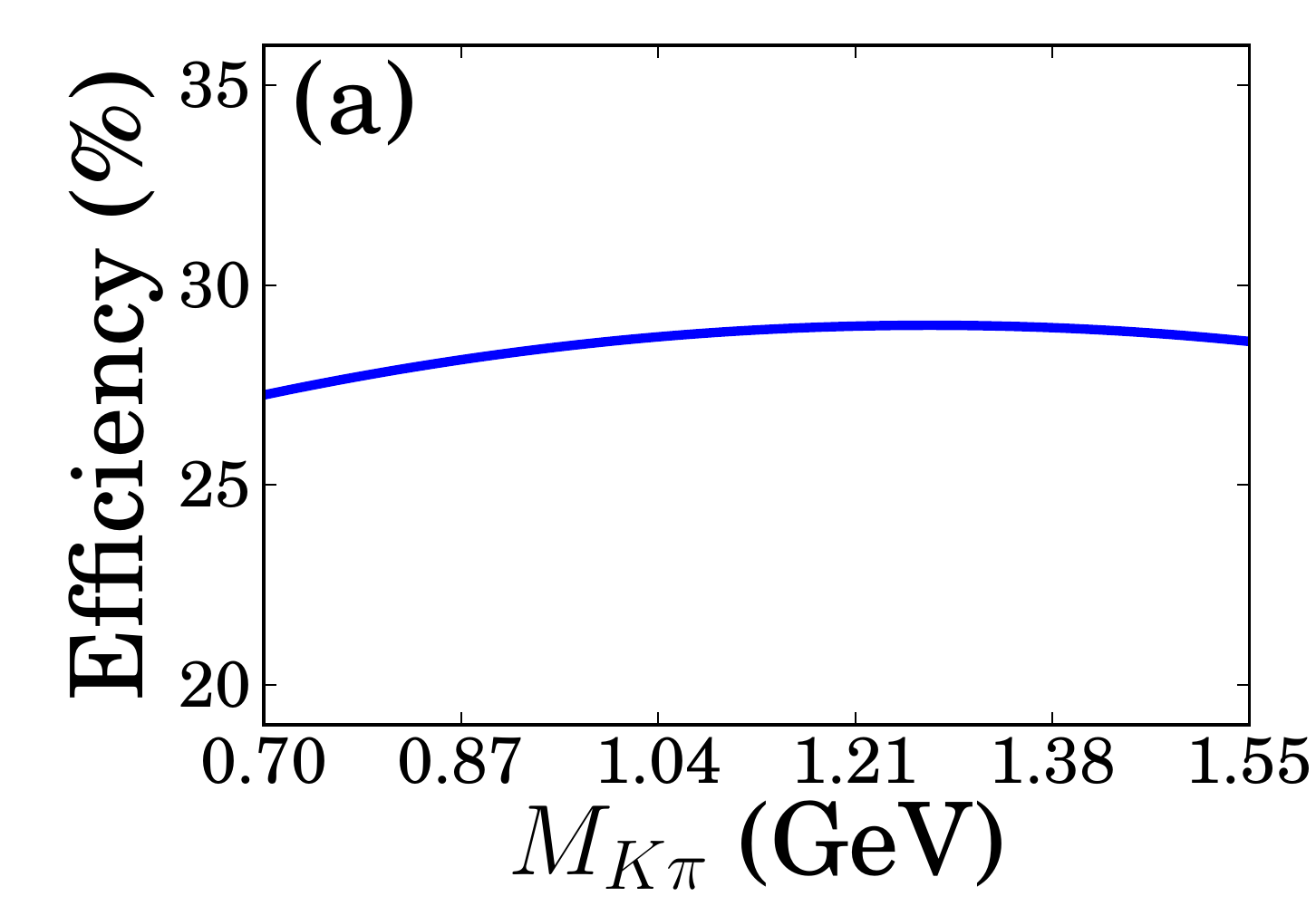}
 \includegraphics[width=0.49\linewidth]{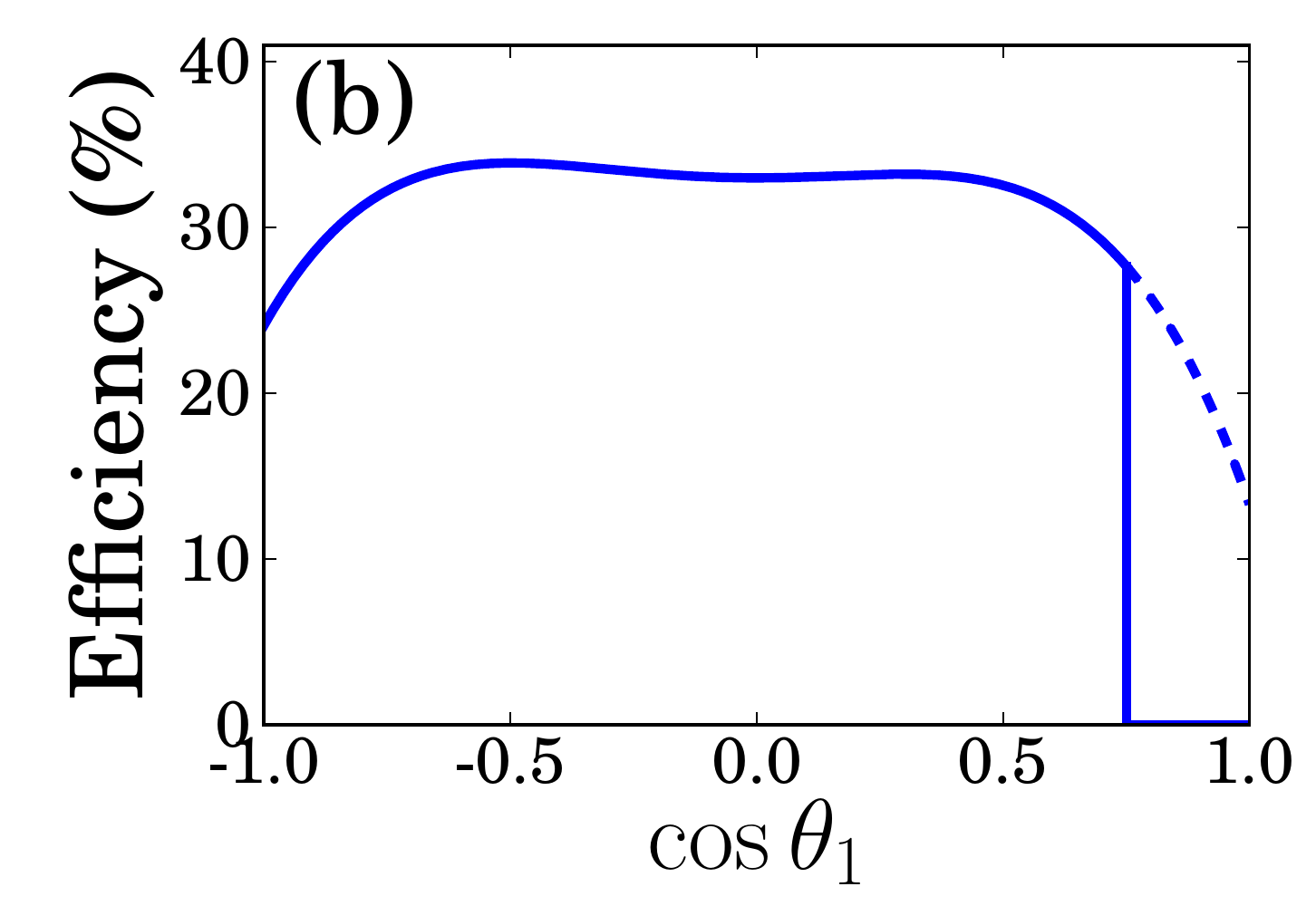}
 \caption{Efficiency as a function of (a) $\mkpi$ and (b) $\cos \theta_1$. In (b) the dashed line indicates the $\cos \theta_1$ region excluded from the analysis.}
\label{fig:efficiency_and_acceptance_plot}
\end{figure}

We derive the four-dimensional efficiency function $\alpha \left( \mkpi, \helthetaone,\helthetatwo,\helphi \right)$ using MC samples of $\BdecayPHSP$ three-body phase space decays. It is found that the efficiency function can be parametrized by the product of one-dimensional projections $\alpha \left( \mkpi, \helthetaone,\helthetatwo,\helphi \right) \equiv \alpha(\mkpi) \times \alpha(\helthetaone) \times \alpha(\helthetatwo) \times \alpha(\helphi)$. We model the efficiency as a function of $\mkpi$ with a second-order polynomial function. The efficiency as a function of $\cos \theta_1$ is parametrized by a fourth-order polynomial function for $\cos \theta_1 < 0.75$ and zero above. Both distributions are shown in Fig.~\ref{fig:efficiency_and_acceptance_plot}. The efficiency as a function of $\cos \theta_2$ and $\Phi$ is found to be uniform.

For a $\BdecayPHSP$ three-body phase-space decay, we obtain an averaged reconstruction efficiency of about $28\%$ within the analysis region. The reconstruction efficiency $\epsilon_{\text{reco},J}$ for a given partial wave $J$ depends on the observed angular distribution and can be obtained only after the polarization is measured. For the partial wave amplitudes with spin $J$ in Eqs.~(\ref{eqn:Swave}) to (\ref{eqn:Dwave}), we compute $\epsilon_{\text{reco},J}$ using
\begin{equation}
\label{eqn:efficiency_per_resonance}
\epsilon_{\text{reco},J} = \frac{\int \alpha \left( \mkpi, \helthetaone,\helthetatwo,\helphi \right) \vert \mathcal{A}_J \vert^2 }{\int \vert \mathcal{A}_J \vert^2} = \frac{n}{d}.
\end{equation}

The numerator $n$ is the integral over the phase space with the efficiency included and is given by
\begin{equation}
n = \int\limits_{m_{K}+m_{\pi}}^{m_{B^0}-m_{\phi}} \int\limits_{-1}^{+1} \int\limits_{-1}^{+1} \int\limits_{-\pi}^{\pi} \alpha \vert \mathcal{A}_J \vert^2 d\mkpi d\cos\theta_1 d\cos\theta_2 d\Phi,
\end{equation}
where $m_{B^0}$, $m_\phi$, $m_K$ and $m_\pi$ are the nominal particle masses that limit the $\mkpi$ phase space. We omit the explicit dependencies of $\alpha$ and $\mathcal{A}_J$ for readability. The denominator of Eq.~(\ref{eqn:efficiency_per_resonance}), $d$, is given by the integral over the full phase space with a uniform efficiency
\begin{equation}
d = \int\limits_{m_{K}+m_{\pi}}^{m_{B^0}-m_{\phi}} \int\limits_{-1}^{+1} \int\limits_{-1}^{+1} \int\limits_{-\pi}^{\pi} \vert \mathcal{A}_J \vert^2 d\mkpi d\cos\theta_1 d\cos\theta_2 d\Phi.
\end{equation}

\section{Partial wave analysis}
We use an unbinned extended maximum-likelihood (ML) fit to extract the 26 parameters related to polarization and $CP$ violation defined in Eqs.~(\ref{eqn:fit_parameters}) and (\ref{eqn:fit_parameters_end}), and denoted $\vec{\mu}$ in the following. The log-likelihood function is given by 
\begin{equation}
\ln \mathcal{L} = \sum_{j = 1}^{N} \ln \left\lbrace \sum_{i = 1}^{N_c} N_i \mathcal{P}_i(\vec{x}_j; \vec{\mu}; \vec{\vartheta}) \right\rbrace - \sum_{i = 1}^{N_c} N_i,
\end{equation}
where $N$ is the total number of candidate events in the data set, $N_c$ is the number of contributions, $N_i$ is the expected number of events for the $i$\textsuperscript{th} contribution, $\mathcal{P}_i$ is the probability density function (PDF) for the $i$\textsuperscript{th} contribution, $\vec{x}_j$ is the nine-dimensional vector of observables for the $j$\textsuperscript{th} event, and $\vec{\vartheta}$ denotes remaining parameters such as those related to PDF shapes.

We include three contributions in our fit model: the signal decay $\BdecayK$ ($i = 1$), peaking background from $\Bdecayf$ decays ($i = 2$), and combinatorial background ($i = 3$). Each event is characterized by a nine-dimensional set of observables $\vec{x}_j = \left\lbrace \mbc, \deltae, \cnb, \mkk, \mkpi, \helthetaone, \helthetatwo, \helphi, Q \right\rbrace$, with the beam-energy-constrained mass $\mbc$, the energy difference $\deltae$, the transformed continuum NN output $\cnb = \ln\left( C_{\rm NB}/(1-C_{\rm NB}) \right)$, the invariant mass of the $\phi$ candidate $\mkk$, the invariant mass of the $K^*$ candidate $\mkpi$, the three helicity angles $\helthetaone$, $\helthetatwo$ and $\helphi$, and the charge $Q=\pm 1$ of the primary kaon from the $B$ meson, denoting the $B$ meson flavor. The transformed $\cnb$ is used instead of $C_{\rm NB}$ as it has a Gaussian-like shape and can be described by an analytic parametrization.

\subsection{PDF parametrization}
The PDF $\mathcal{P}_i(\vec{x}_j; \vec{\mu}; \vec{\vartheta})$ for a given contribution $i$ is constructed as a joint PDF of the distributions of the observables $\vec{x}$. With a few exceptions, explained below, we find no significant correlations among the fit observables. We use the method described in Ref.~\cite{CAT} to check for linear and non-linear correlations among the observables using MC samples as well as sideband and off-resonance data for cross-checks.

The signal PDF for $\BdecayK$ is modeled with a double Gaussian function for $\mbc$. The $\deltae$ distribution is modeled with the sum of a Gaussian and two asymmetric Gaussian functions. In addition, to take into account a significant linear correlation between $\mbc$ and $\deltae$ for the signal, the mean of the $\deltae$ distribution is parametrized by a linear function of $\mbc$. The $\cnb$ distribution is parametrized by a sum of two asymmetric Gaussian functions. The $\phi$ candidate mass $\mkk$ is modeled by a relativistic spin-dependent BW convolved with a Gaussian function to account for resolution effects; the BW parameters can be found in Table~\ref{tab:resonance_parameters_KK}. For $\mkpi$, the helicity angles and $Q$ we refer to Eq.~(\ref{eqn:width_signal}), which we multiply with the experimentally derived efficiency function $\alpha \left( \mkpi, \helthetaone,\helthetatwo,\helphi \right)$ to obtain the mass-angular signal PDF.

\begin{table}[ht]
\caption{Parameters used for the $\phi$ resonance are taken from~\cite{PDG}, except for $r$, we make an assumption based on the values found in $K\pi$ scattering. For $f_0(980)$, we use values from BES~\cite{BES}.}
\label{tab:resonance_parameters_KK}
\begin{tabular}{ccc} \hline \hline
 & $\phi$ & $f_0(980)$ \\
Parameter & $J=1$  & $J=0$\\ \hline
$m_J$ (MeV) & $1019.455 \pm 0.020$ & $965 \pm 10$ \\
$\Gamma_J$ (MeV) & $4.26 \pm 0.04$ & $\cdots$ \\
$r$ (GeV$^{-1}$) & $3.0 \pm 1.0$ & $\cdots$\\
$g_\pi$ (MeV) & $\cdots$ & $165 \pm 18$ \\
$g_K$ (MeV) & $\cdots$ & $(4.21\pm0.33)g_\pi$ \\
\hline \hline 
\end{tabular}
\end{table}

The peaking background PDF for $\Bdecayf$ is constructed using the same parametrization as signal for $\mbc$, $\deltae$ and $\cnb$. The $\mkk$ distribution of the $f_0(980)$ candidates is modelled by a Flatt\'e function~\cite{F76}. The resonance parameters are given in Table~\ref{tab:resonance_parameters_KK}. The $\mkpi$ distribution is parametrized by a relativistic spin-dependent BW for $\Pwave$ using the same parameters as the signal component. The angular distribution of this pseudoscalar to scalar--vector decay is uniform in $\helthetatwo$ and $\helphi$, and is proportional to $\helthetaonesquared$; we correct for detector acceptance effects. We use a distribution with equal probability for the two values of $Q$.

The combinatorial background PDF follows an empirically determined shape for the $\mbc$ distribution, given by
\begin{equation}
f(\mbc) \propto \mbc \sqrt{1-\frac{\mbc^2}{E_{\rm beam}^{*2}} } \exp \left[ c \left(1-\frac{\mbc^2}{E_{\rm beam}^{*2}} \right) \right],
\end{equation}
where $c$ is a free parameter. This function was first introduced by the ARGUS Collaboration~\cite{Argus}. The $\deltae$ distribution is parametrized by a first-order polynomial function. The $\cnb$ distribution is parametrized with a sum of two asymmetric Gaussians. To account for background that contains real $\phi$ candidates and a non-resonant component, the $\mkk$ distribution is parametrized by the sum of resonant and non-resonant contributions. Similar to signal, the resonant contribution is parametrized with a relativistic spin-dependent BW convolved with the same resolution function. The non-resonant component is described by a threshold function as
\begin{equation}
f(\mkk) \propto \arctan \left( \sqrt{(\mkk - 2 m_K)/a} \right),
\end{equation}
where $m_K$ is the $K^\pm$ mass and $a$ a free parameter in the fit. The $\mkpi$ distribution is also parametrized by a sum of resonant and non-resonant components. The resonant component from $\Pwave$ is modelled with a relativistic spin-dependent BW using the same parameters as the signal component. The non-resonant contribution is parametrized by a fourth-order Chebyshev polynomial. We find a significant non-linear correlation between $\mkpi$ and $\helthetaone$ in the non-resonant component of the combinatorial background. The resonant component in $\mkpi$ is uniform in $\helthetaone$, whereas the non-resonant contribution is parametrized by a fifth-order Chebyshev polynomial, where the parameters depend linearly on $\mkpi$. The $\helthetatwo$ distribution is parametrized by a second-order Chebyshev polynomial and the distributions in $\helphi$ and $Q$ are uniform. The combinatorial background PDF is verified using off-resonance and sideband data. The $2\%$ contribution due to the combinatorial background from $\bbar$ events, which is present in the sideband, has no significant effect on the shape parameters.

We use sideband data events to determine the free parameters of the combinatorial background PDF. Due to the presence of a clear $\phi$ peak in these events, we also determine the $\mkk$ resolution (about 1~MeV) from this fit and use it for the signal model in the nominal fit region.

The $\mbc$, $\deltae$ and $\cnb$ distributions of the signal and peaking background components are cross-checked by fitting to a large-statistics control sample of $\BdecayControl$ events. In the control channel, we find excellent agreement between data and simulations for the distributions of $\mbc$ and $\cnb$. We also confirm the linear correlation between $\mbc$ and $\deltae$ and that our conditional PDF based on MC simulations describes the data well. The only difference we observe is due to the $\deltae$ resolution, for which we derive a scale factor $s = 1.124 \pm 0.062$ by comparing data and MC events in the control sample. The scale factor is applied to the signal and peaking background model on data.

In the fit to the nominal fit region, we use the combinatorial background model derived from the fit to the sideband data and fix all parameters except one. The parameter $c$ from the $\mbc$ background shape is the only floated parameter related to the combinatorial background in the nominal fit region, as it is sensitive to the shape towards the kinematic endpoint of the $\mbc$ distribution and can be determined from the sideband only with large uncertainties. Beside this, we float the three yields $N_i$ and the 26 signal parameters $\vec{\mu}$.

The log-likelihood function is minimized using the MINUIT~\cite{Minuit} algorithm in the RooFit package~\cite{RooFit} within the ROOT framework~\cite{ROOT}. The RooFit package provides an interface that allows us to extend it with the PDFs necessary for the partial wave analysis. Further, it provides functionality for the normalization of PDFs and visualization of fit results.

\subsection{Optimization of normalization integrals}
The normalization integrals in Eq.~(\ref{eqn:normalization}) require four-dimensional numeric integration, which is computationally expensive. As the weights in $\mathcal{M}$ are adjusted during an ML fit, this operation needs to be performed thousands of times. However, such integration can be optimized drastically when certain conditions are satisfied.

The integration over a simple matrix element $\vert\mathcal{M}\vert^2$ with two amplitudes $\mathcal{A}_i(\vec{x})$ ($i=0,1$) depending on observables $\vec{x}$ and their complex weights $A_i = a_i \exp (i \varphi_i)$
\begin{equation}
\int \vert \mathcal{M} \vert^2 d\vec{x} = \int \vert A_0 \cdot \mathcal{A}_0(\vec{x}) + A_1 \cdot \mathcal{A}_1(\vec{x}) \vert^2 d\vec{x},
\end{equation}
can be expanded to
\begin{equation}
\begin{split}
\int \vert \mathcal{M} \vert^2 d\vec{x} =\ & a_0^2 \int \vert \mathcal{A}_0(\vec{x}) \vert^2 d\vec{x} + a_1^2 \int \vert \mathcal{A}_1(\vec{x}) \vert^2 d\vec{x}\\
 +& 2 a_0 a_1 \cos \Delta\varphi \int \operatorname{Re}\lbrace \mathcal{A}_0(\vec{x}) \mathcal{A}_1^*(\vec{x}) \rbrace d\vec{x}\\
 -& 2 a_0 a_1 \sin \Delta\varphi \int \operatorname{Im} \lbrace \mathcal{A}_0(\vec{x}) \mathcal{A}_1^*(\vec{x}) \rbrace d\vec{x}, 
\end{split}
\end{equation}
with $\Delta \varphi = \varphi_0 - \varphi_1$. Given $n$ amplitudes $\mathcal{A}_i$, we always obtain $n$ integrals over $\mathcal{A}_i$ squared and $(n^2-n)/2$ integrals over the real and imaginary parts of the product of two amplitudes, respectively. If the amplitudes $\mathcal{A}_i$ have no free parameters, all integrals become constant as only the weights are adjusted.

In the context of this analysis, we have $n=7$ helicity amplitudes, resulting in $49$ constant integrals, as parameters such as resonance masses, interaction radii and other similar quantities are fixed. These integrals are computed once with high precision and are then used on demand, thereby significantly reducing the amount of CPU time. This method is several orders of magnitude faster than the numeric integration in each iteration of the fit. For cross-checks, we performed a comparison between our approach and the numeric integration. This exercise confirmed the validity of our approach but required several days of CPU time.

This technique can also be used to improve the computation of projection integrals onto one dimension $d$ in $\vec{x}$ for a fixed value of $x_d$. In a typical projection plot, hundreds of $(\dim{\vec{x}}-1)$-dimensional integrations that could require several hours of CPU time are necessary per plot. These can be computed in parallel and stored on a large scale cluster. If loaded on demand, the improvement in speed is again several orders of magnitude.

\subsection{Validation}
The entire analysis is performed as a blind analysis and all methods are tested and fixed before being applied to the data. Beside the above mentioned cross-checks of PDFs on the control channel, sideband and off-resonace data, and the optimized computation of normalization integrals, we have made further studies to validate the analysis chain.

The $b \to c$ and $b \to s$ MC samples are used to search for possible backgrounds beside the already described $\Bdecayf$ and $\BdecayDsK$. In the case of particle misidentification, $\BdecayPhiPhi$ or $\BdecayPhiRho$ (for example) could mimic signal but no statistically significant contribution is expected from these or other decays. After unblinding the data, we create projections of the data and likelihood function in regions that would show an enhancement from any peaking decays present in the data sample. In all cases, the difference between data and the fit model is within the statistical uncertainty.

We also generate a large number of independent pseudo-experiments from the PDF, using results from previous measurements on the polarization parameters. We also use pseudo-experiments with different assumptions about the polarization or the level of additional $CP$ violation. In all pseudo-experiments, we find the fit procedure to be robust and the mean and width of the pull distributions to be consistent with the expectation.

We further use simulated $\BdecayPHSP$ phase space signal events, which we reweight according to different polarizations, as well as the four independent samples of $b \to c$ and $q\bar{q}$ events to perform fits. Again, the expected inputs are reproduced within the statistical errors.

Finally, we check the fit stability with respect to multiple solutions by fitting samples repeatedly with random starting values for the signal parameters. The MINUIT algorithm is not always able to find the global minimum for a given sample and only in about $30\%$ of fits the correct minimum is found. In the remaining $70\%$, the algorithm is trapped in a local minimum, which has significantly larger $-2\ln \mathcal{L}$ value ($-2\Delta\ln\mathcal{L} > 50$) than the global minimum, or stops without finding a minimum. We therefore repeat the final fit on data 100 times and select the best solution by the lowest negative log-likelihood value. Using 100 repetitions, we never observed the global minimum to be not found in the pseudo-experiments; the lowest fraction observed has been around $25\%$.

\subsection{Systematics}
\label{sec:systematics}
We split various sources of systematics into two main groups. Systematics in the first group are summarized in Table~\ref{tab:systematics_error_efficiency} and include uncertainties that enter the computation of the branching fraction and are rather decoupled from the polarization and $CP$ violation parameters. Systematics from the second group are summarized in Tables~\ref{tab:systematics_error_triple_product_correlations} and \ref{tab:systematics_error_physics_parameters}, and include uncertainties that affect the polarization and $CP$ violation parameters, including triple-product correlations.

Due to uncertainties in the reconstruction efficiency for charged tracks, we assign $0.35\%$ uncertainty per track, which results in $1.4\%$ total uncertainty. These values have been estimated from a study of partially reconstructed $D^{*+} \to D^0 \pi^+ \to (K_S^0 \pi^+ \pi^-) \pi^+$ decays. The uncertainty due to PID selection is estimated from $D^{*+} \to D^0 \pi^+ \to (K^- \pi^+) \pi^+$ samples and tabulated as a function of track momentum and polar angle. The assigned value of $(3.3-3.4)\%$ varies among different partial waves as it depends slightly on the polarization. To evaluate the difference between data and simulations for the selection based on $C_{\rm NB}$, we use the $\BdecayControl$ control sample but find no need for any correction. We assign a systematic uncertainty of $0.7\%$ on the $C_{\rm NB}$ requirement due to finite data statistics in the control channel. Due to limited MC statistics, we assign $0.5\%$ uncertainty on the absolute scale of efficiency. Uncertainties due to daughter branching fractions of $\phi$ and $\Dwave$ are taken from Ref.~\cite{PDG}. Finally, the uncertainty on the total number of $\bbar$ pairs in data is estimated to be $1.4\%$. All uncertainties are summarized in Table~\ref{tab:systematics_error_efficiency}, including the total uncertainty estimated by adding the individual errors in quadrature.

\begin{table}[ht]
\caption{Systematic errors ($\%$) that enter only the calculation of the branching fraction.}
\label{tab:systematics_error_efficiency}
\begin{tabular}{lccc} \hline \hline
 & $\phi \Swave$ & $\phi \Pwave$ & $\phi \Dwave$ \\
 & $J=0$ & $J=1$ & $J=2$ \\ \hline
Track reconstruction & $1.4$ & $1.4$ & $1.4$\\
PID selection & $3.3$ & $3.3$ & $3.4$\\
$C_{\rm NB}$ requirement & $0.7$ & $0.7$ & $0.7$\\
MC statistics & $0.5$ & $0.5$ & $0.5$ \\
$\phi$ branching fraction & $1.0$ & $1.0$ & $1.0$ \\
$K_2^*$ branching fraction & $\cdots$ & $\cdots$ & $2.4$ \\
$N_{\bbar}$ & $1.4$ & $1.4$ & $1.4$ \\ \hline
Total & $4.1$ & $4.1$ & $4.8$ \\ 
 \hline\hline
\end{tabular}
\end{table}

\begin{table}[ht]
\caption{Systematic errors (absolute values) on the triple-product correlations for $\BdecayP$. See Table~\ref{tab:systematics_error_physics_parameters} for column details.}
\label{tab:systematics_error_triple_product_correlations}
\begin{tabular}{lcccccc} \hline\hline
Parameter & PDF & Eff. & SCF & $KK$ & Interf. & Total \\ \hline
               $A_{TB^0}^{0}$ & $0.003$ & $0.003$ & $0.003$ & $0.002$ & $0.008$ & $0.010$ \\
       $A_{TB^0}^{\parallel}$ & $0.003$ & $0.003$ & $0.003$ & $0.001$ & $0.003$ & $0.006$ \\
              $A_{T\bar{B}^0}^{0}$ & $0.004$ & $0.001$ & $0.006$ & $0.000$ & $0.012$ & $0.014$ \\
      $A_{T\bar{B}^0}^{\parallel}$ & $0.002$ & $0.002$ & $0.003$ & $0.000$ & $0.010$ & $0.011$ \\
        $\mathcal{A}_{T}^{0}$ & $0.009$ & $0.006$ & $0.014$ & $0.003$ & $0.015$ & $0.023$ \\
$\mathcal{A}_{T}^{\parallel}$ & $0.087$ & $0.061$ & $0.511$ & $0.012$ & $0.004$ & $0.522$ \\
\hline\hline
\end{tabular}
\end{table}

\begin{table*}[ht]
\caption{Systematic errors (absolute values) on the physics parameters defined in Table~\ref{tab:physics_parameter_definition}. The fit fractions per partial wave $FF_J$ are defined in Sec.~\ref{sec:results}. In addition, we show the relative errors on parameters that enter the calculation of the branching fraction. The uncertainties are due to PDF parametrization, efficiency function, SCF, uncertainties on the $KK$ shape, $KK$ interference effects and charge asymmetry in the reconstruction.}
\label{tab:systematics_error_physics_parameters}
\begin{tabular}{lccccccc} \hline\hline
                 Parameter & PDF     & Eff. & SCF     & $KK$    & Interf. & Charge   & Total \\ \hline
                 $N_{\rm sig}$ &$25.8$ & $1.4$ & $2.9$ &$10.7$ & $0.8$ & $\cdots$ & $28.1$ \\
                    $FF_0$ & $0.021$ & $0.002$ & $0.002$ & $0.003$ & $0.002$ & $\cdots$ & $0.021$ \\
       $\mathcal{A}_{CP0}$ & $0.008$ & $0.003$ & $0.006$ & $0.001$ & $0.005$ & $0.012$  & $0.017$ \\
                    $FF_1$ & $0.013$ & $0.007$ & $0.001$ & $0.004$ & $0.002$ & $\cdots$ & $0.015$ \\
       $\mathcal{A}_{CP1}$ & $0.004$ & $0.002$ & $0.003$ & $0.002$ & $0.016$ & $0.012$  & $0.021$ \\
                    $FF_2$ & $0.017$ & $0.005$ & $0.001$ & $0.001$ & $0.001$ & $\cdots$ & $0.018$ \\
       $\mathcal{A}_{CP2}$ & $0.025$ & $0.012$ & $0.013$ & $0.001$ & $0.000$ & $0.012$  & $0.033$ \\
                  $f_{L1}$ & $0.005$ & $0.016$ & $0.002$ & $0.005$ & $0.002$ & $\cdots$ & $0.018$ \\
             $f_{\perp 1}$ & $0.003$ & $0.007$ & $0.001$ & $0.003$ & $0.001$ & $\cdots$ & $0.008$ \\
      $\phi_{\parallel 1}$ & $0.015$ & $0.005$ & $0.009$ & $0.002$ & $0.010$ & $\cdots$ & $0.020$ \\
          $\phi_{\perp 1}$ & $0.014$ & $0.005$ & $0.013$ & $0.004$ & $0.037$ & $\cdots$ & $0.042$ \\
             $\delta_{01}$ & $0.078$ & $0.018$ & $0.006$ & $0.007$ & $0.011$ & $\cdots$ & $0.081$ \\
     $\mathcal{A}_{CP1}^0$ & $0.003$ & $0.002$ & $0.003$ & $0.001$ & $0.005$ & $\cdots$ & $0.007$ \\
 $\mathcal{A}_{CP1}^\perp$ & $0.006$ & $0.004$ & $0.004$ & $0.001$ & $0.008$ & $\cdots$ & $0.011$ \\
$\Delta\phi_{\parallel 1}$ & $0.009$ & $0.004$ & $0.005$ & $0.001$ & $0.005$ & $\cdots$ & $0.012$ \\
    $\Delta\phi_{\perp 1}$ & $0.008$ & $0.005$ & $0.012$ & $0.001$ & $0.010$ & $\cdots$ & $0.018$ \\
       $\Delta\delta_{01}$ & $0.006$ & $0.004$ & $0.006$ & $0.001$ & $0.001$ & $\cdots$ & $0.010$ \\
                  $f_{L2}$ & $0.011$ & $0.006$ & $0.003$ & $0.000$ & $0.000$ & $\cdots$ & $0.012$ \\
             $f_{\perp 2}$ & $0.008$ & $0.003$ & $0.003$ & $0.001$ & $0.000$ & $\cdots$ & $0.009$ \\
      $\phi_{\parallel 2}$ & $0.138$ & $0.072$ & $1.314$ & $0.009$ & $0.017$ & $\cdots$ & $1.323$ \\
          $\phi_{\perp 2}$ & $0.121$ & $0.049$ & $0.010$ & $0.007$ & $0.013$ & $\cdots$ & $0.131$ \\
             $\delta_{02}$ & $0.177$ & $0.053$ & $0.010$ & $0.002$ & $0.002$ & $\cdots$ & $0.185$ \\
     $\mathcal{A}_{CP2}^0$ & $0.008$ & $0.001$ & $0.002$ & $0.000$ & $0.000$ & $\cdots$ & $0.008$ \\
 $\mathcal{A}_{CP2}^\perp$ & $0.077$ & $0.020$ & $0.030$ & $0.010$ & $0.002$ & $\cdots$ & $0.085$ \\
$\Delta\phi_{\parallel 2}$ & $0.254$ & $0.062$ & $0.979$ & $0.010$ & $0.017$ & $\cdots$ & $1.014$ \\
    $\Delta\phi_{\perp 2}$ & $0.101$ & $0.023$ & $0.013$ & $0.006$ & $0.014$ & $\cdots$ & $0.106$ \\
       $\Delta\delta_{02}$ & $0.011$ & $0.003$ & $0.009$ & $0.001$ & $0.003$ & $\cdots$ & $0.015$ \\
\hline
          $N_{\rm sig}$ $(\%)$ & $2.3$ & $0.1$ & $0.3$ & $1.0$ & $0.1$ & $\cdots$ & $2.5$ \\
             $FF_0$ $(\%)$ & $7.7$ & $0.7$ & $0.7$ & $1.1$ & $0.7$ & $\cdots$ & $7.9$ \\
             $FF_1$ $(\%)$ & $2.2$ & $1.2$ & $0.2$ & $0.7$ & $0.3$ & $\cdots$ & $2.6$ \\                                                            	         $FF_2$ $(\%)$ &$17.2$ & $5.1$ & $1.0$ & $1.0$ & $1.0$ & $\cdots$ &$18.0$ \\
\hline\hline 
\end{tabular}
\end{table*}

As for systematics on the polarization and $CP$ violation parameters, we first consider uncertainties due to the PDF model. The external inputs on resonance masses, widths, and other fixed parameters in our measurement are given in Tables~\ref{tab:resonance_parameters_Kpi} and \ref{tab:resonance_parameters_KK} together with their uncertainties. Besides parameters shown in these tables, we have fixed combinatorial background shape parameters from the fit to sideband data as well as signal and peaking background shape parameters from fits to MC and the control channel. We vary all those parameters one by one by $\pm 1 \sigma$, with $\sigma$ being their statistical uncertainty, and add the differences with respect to the nominal fit result in quadrature and assign it as the systematic uncertainty. Most PDF systematics are dominated by the uncertainties on the external inputs.

We also study the effect of neglecting the resolution in $\mkpi$ and the three helicity angles by creating pseudo-experiments, which we fit with and without applying an additional Gaussian smearing with the resolution derived from MC samples. We find the relative difference to be $\mathcal{O}(10^{-4})$ and thus negligible as compared to other systematics.

The uncertainty in the efficiency function is estimated by varying the efficiency model parameters one by one by $\pm 1 \sigma$. The differences between the efficiency functions for $B^{0}$ and $\bar{B}^{0}$ are found to be smaller than the statistical uncertainties. Again, we add the differences to the nominal fit result in quadrature and take it as the systematic uncertainty.

To assess the impact of the remaining fraction of SCF events, we generate pseudo-experiments and fit them with and without adding SCF events from the MC samples. We use SCF events from MC simulations that correspond to the observed polarization. The mean of the residual between fits with and without additional SCF events is found to be consistent with zero and we take the width of the obtained residual distribution as a systematic uncertainty.

Concerning the shape parametrization of the invariant $K^+K^-$ mass $\mkk$ for the peaking background of $\Bdecayf$, we consider also a non-resonant contribution from $\BdecayKK$ events. We modify the fit model to allow for a coherent sum of both components with relative amplitude and phase between them. Taking into account the change in the number of degrees of freedom, negative log-likelihoods obtained from this alternative fit and the nominal fit yield equally good solutions. The coherent sum shows a very strong destructive interference, which is also often observed in Dalitz analyses (e.\,g. Ref.~\cite{KKinterference}) that include $K^+K^-$. We therefore choose the nominal fit model as the default model and take the difference between the two fits as a systematic uncertainty. We do not consider a model with only $\BdecayKK$, as this hypothesis shows significant deviations between data and fit model in the $\mkk$ region below the $\phi$ peak.

\begin{figure*}[ht!]
\centering
	\includegraphics[width=0.245\linewidth]{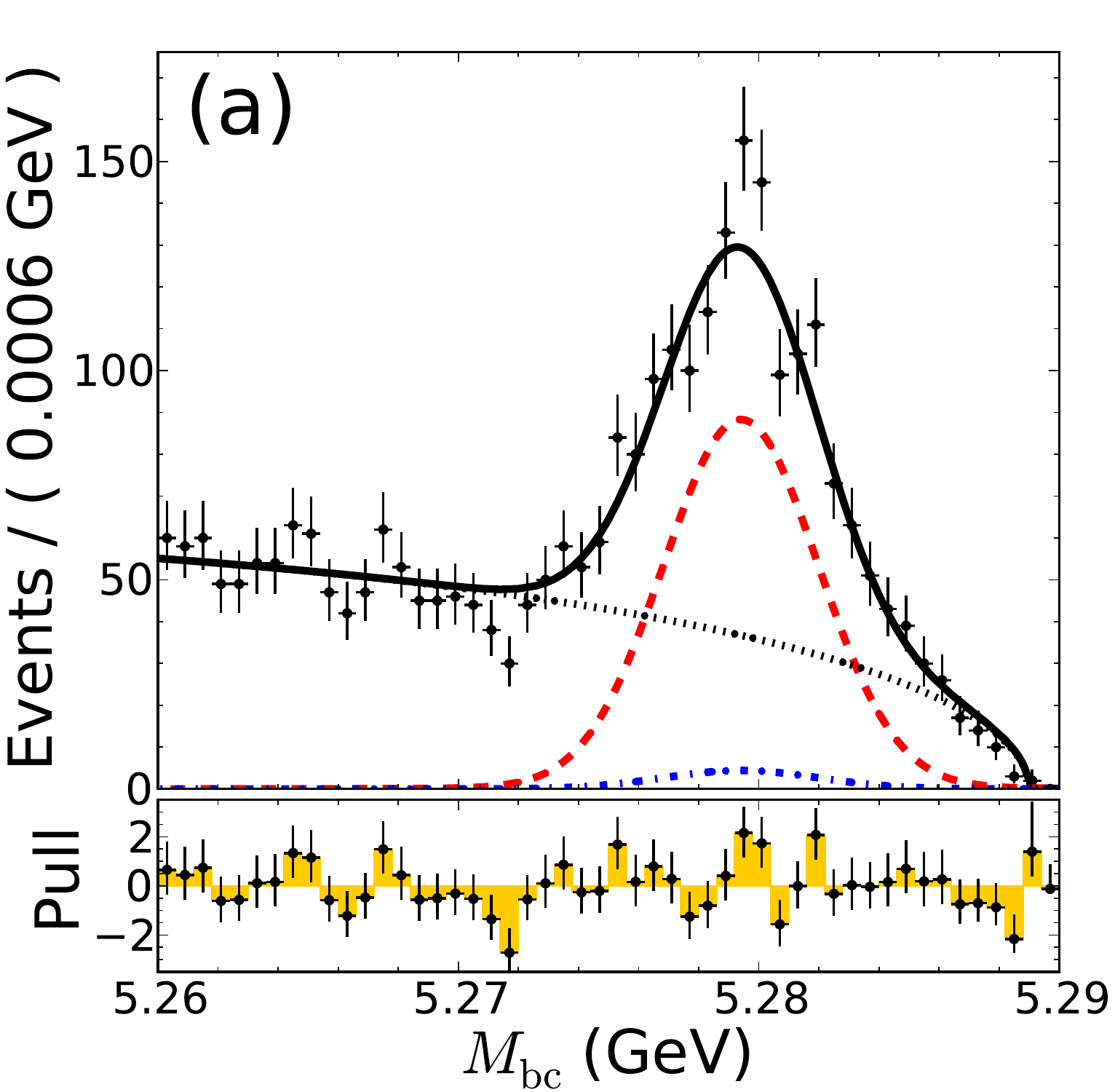}
	\includegraphics[width=0.245\linewidth]{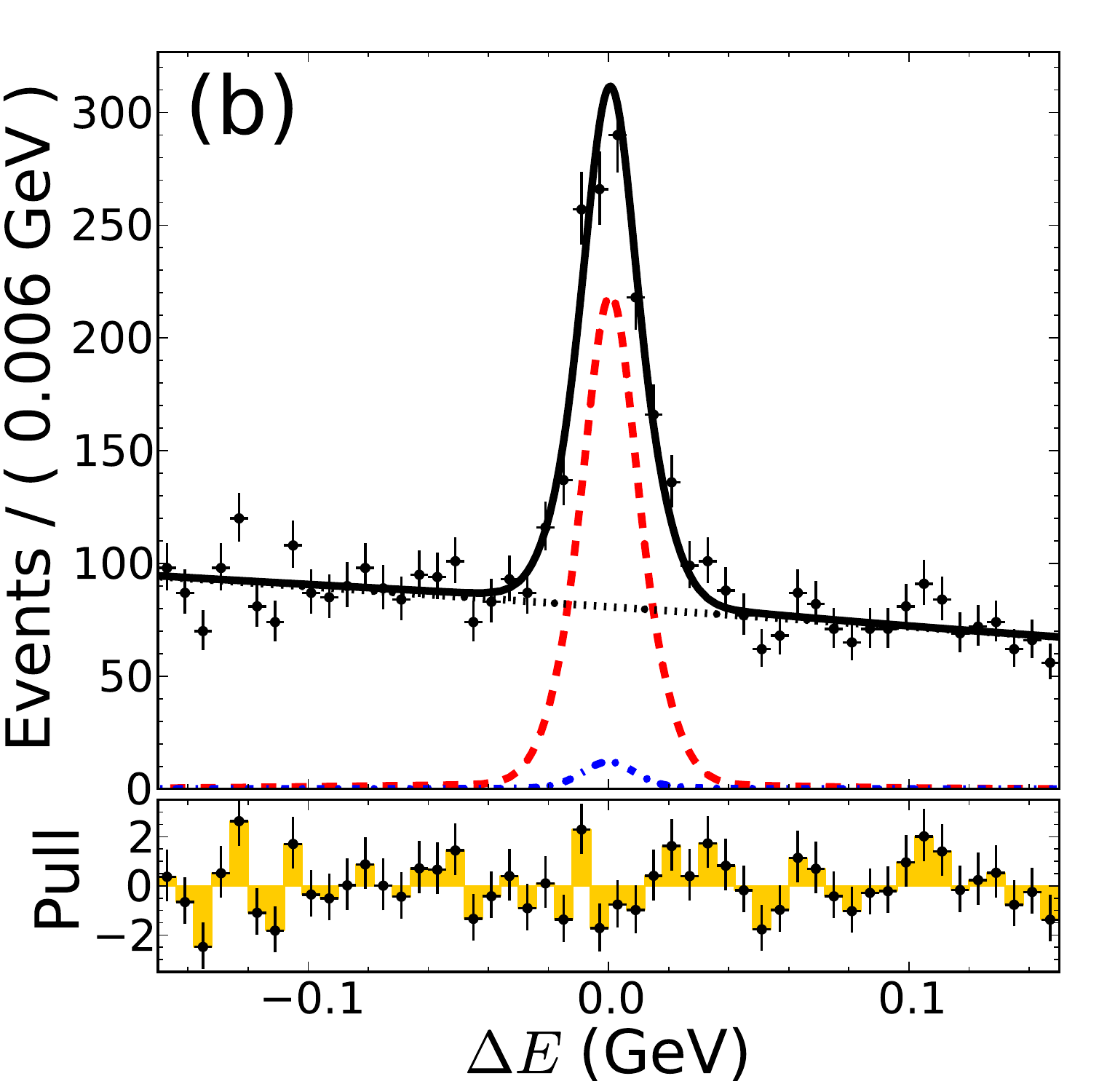}
	\includegraphics[width=0.245\linewidth]{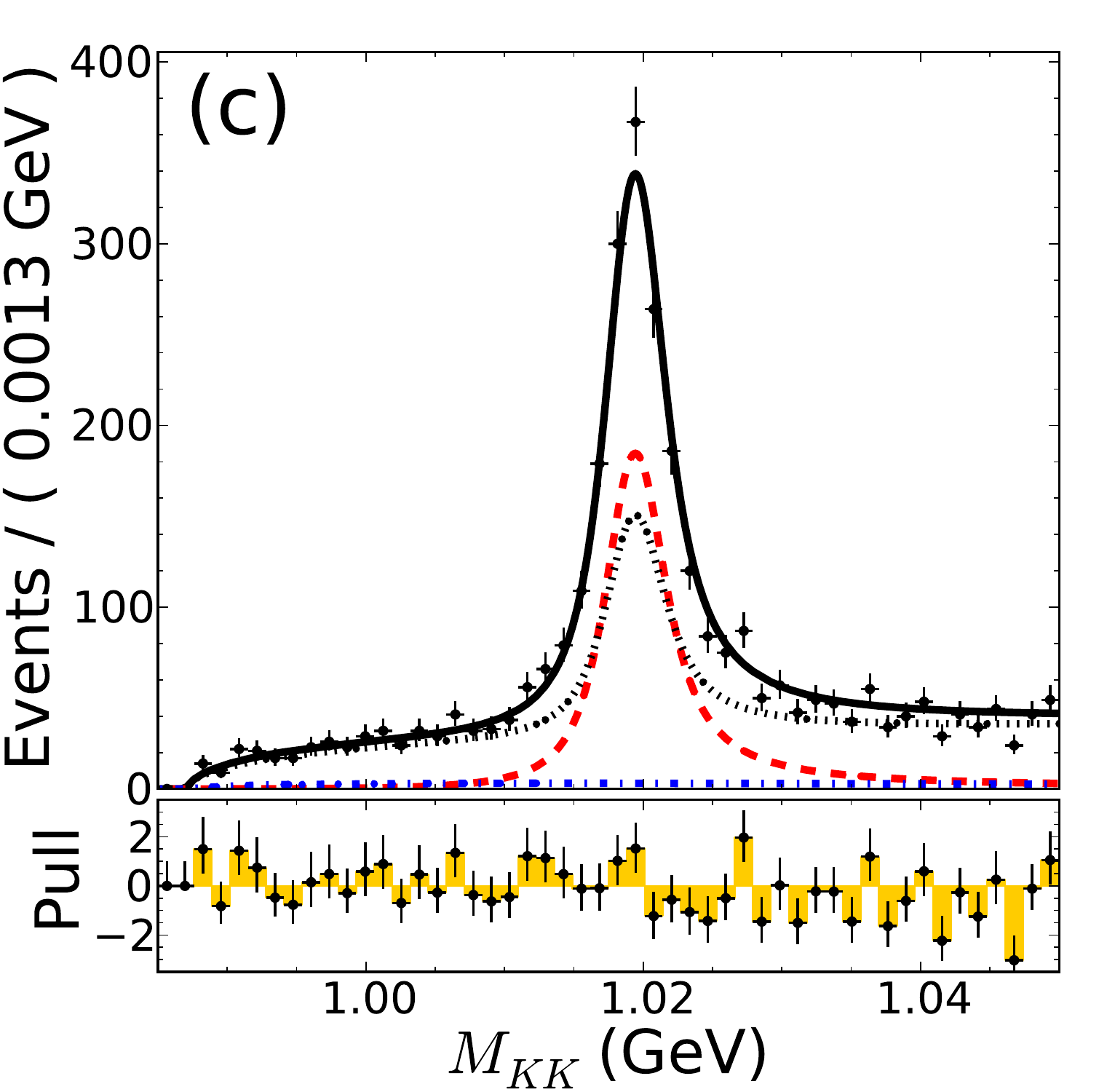}
	\includegraphics[width=0.245\linewidth]{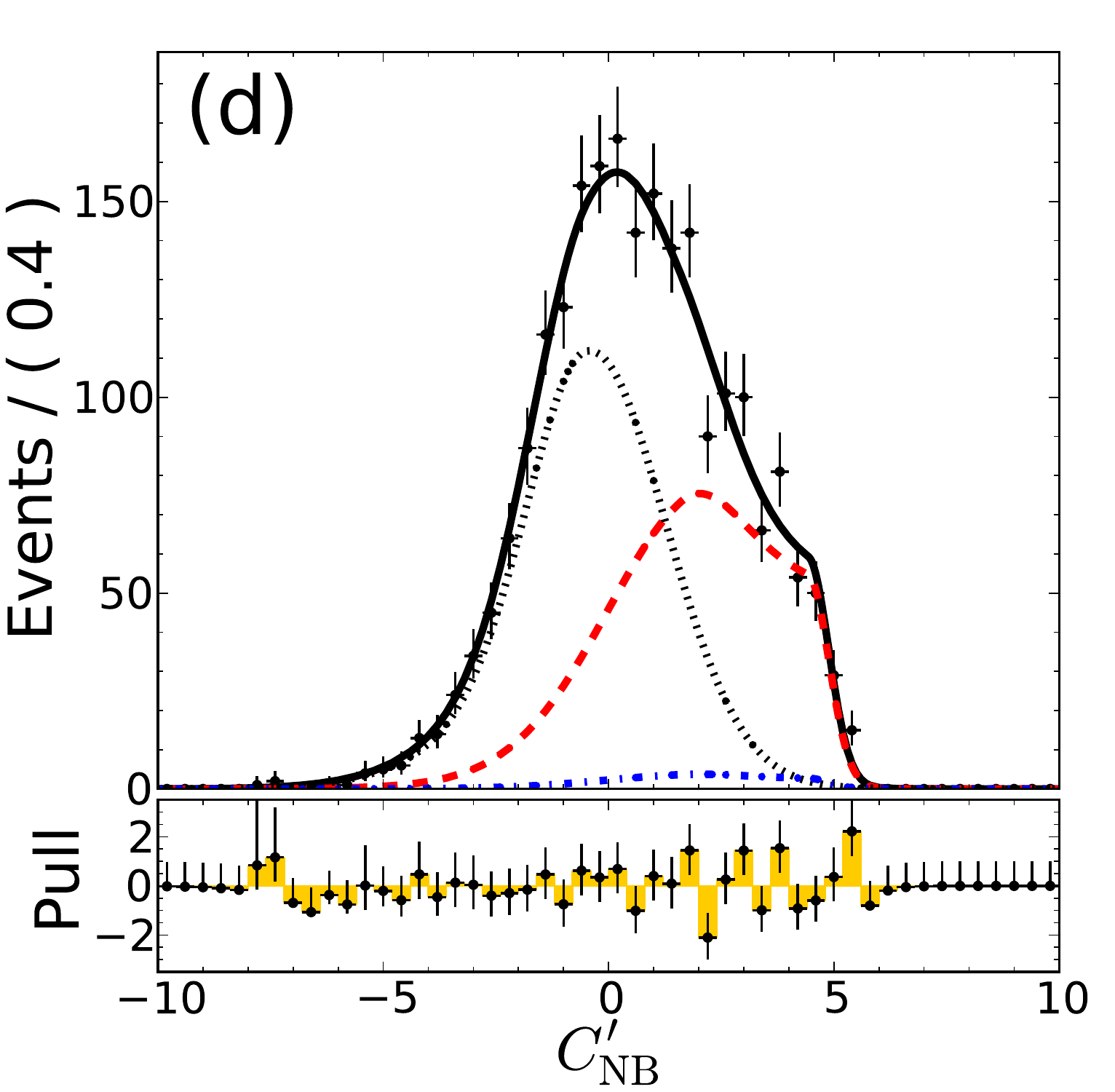}
	\caption{Projections onto the observable (a) $\mbc$, (b) $\deltae$, (c) $\mkk$, and (d) $\cnb$ for $\Bdecay$ and $\BdecayCC$ combined. The data distributions are shown by black markers with error bars whereas the overall fit function, combinatorial background, signal and peaking background are shown with solid black, dotted black, dashed red and dash-dotted blue curves, respectively. For each projection, the data points and fit projections are shown after a signal-enhancing selection (see text) on the other three observables; e.\,g., in (a) a requirement on (b),(c) and (d) is applied.}
	\label{fig:DATA_projections_signal_range}
\end{figure*}

\begin{figure*}[ht!]
\centering
	\includegraphics[width=0.245\linewidth]{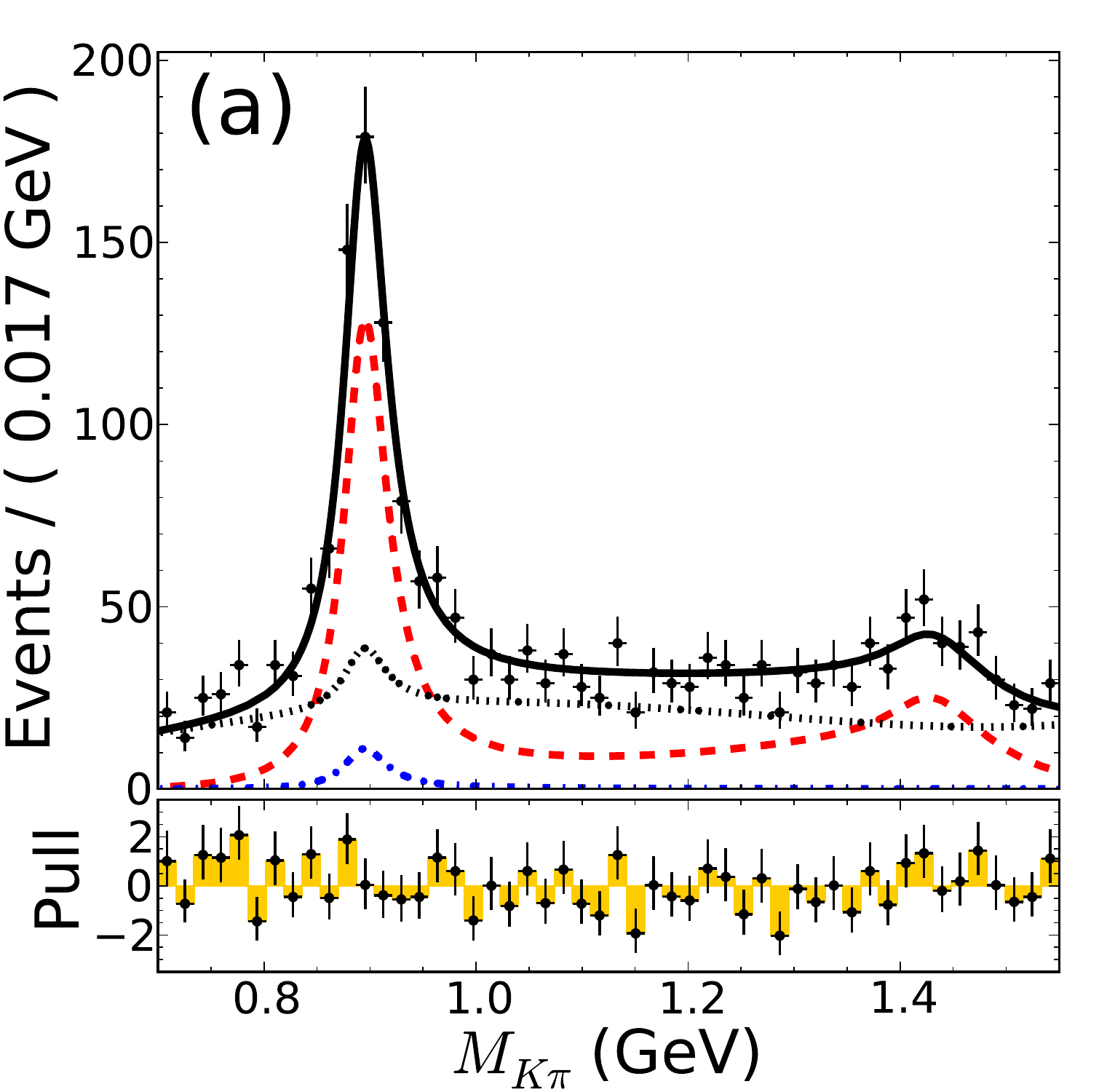}
	\includegraphics[width=0.245\linewidth]{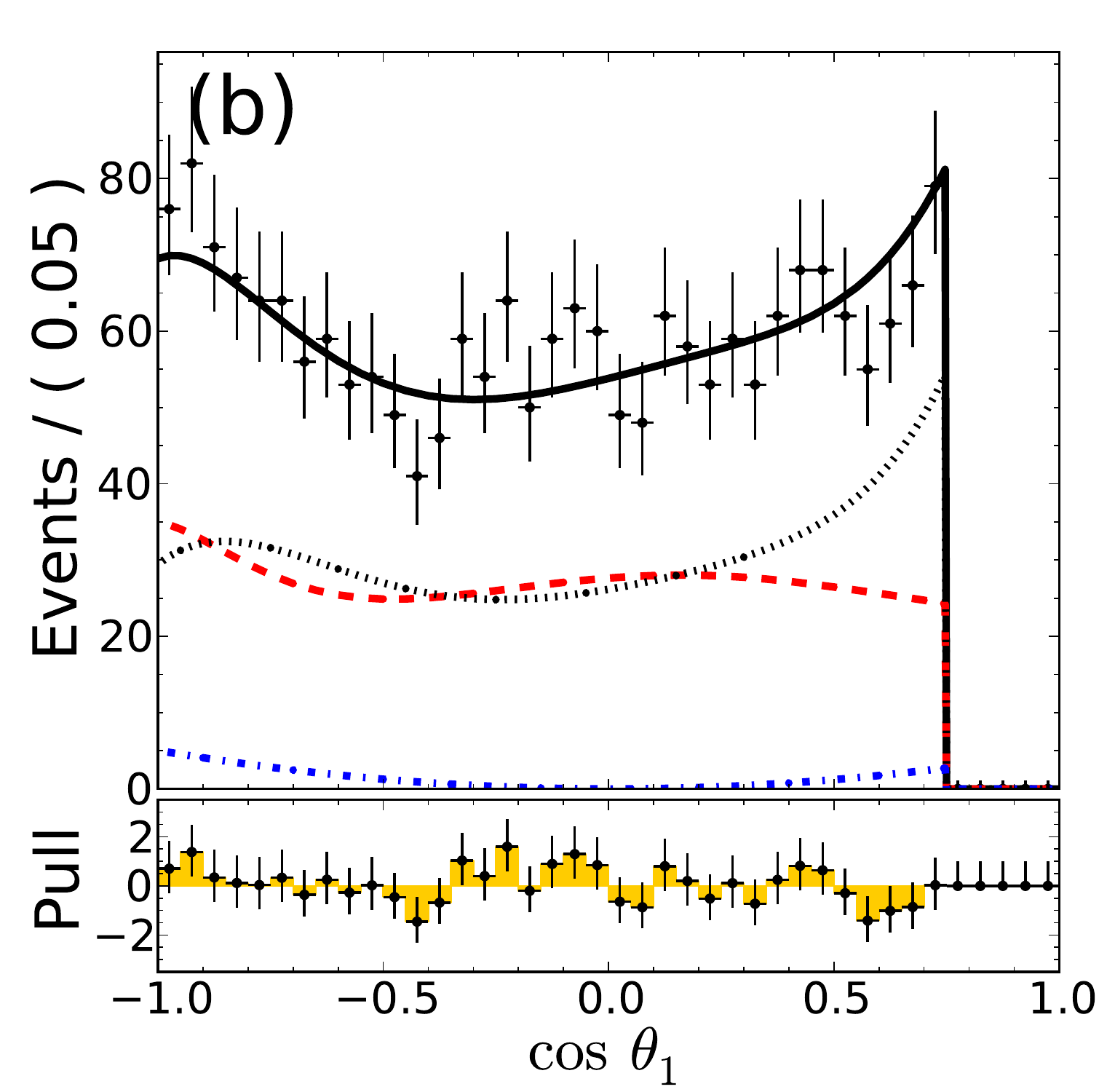}
	\includegraphics[width=0.245\linewidth]{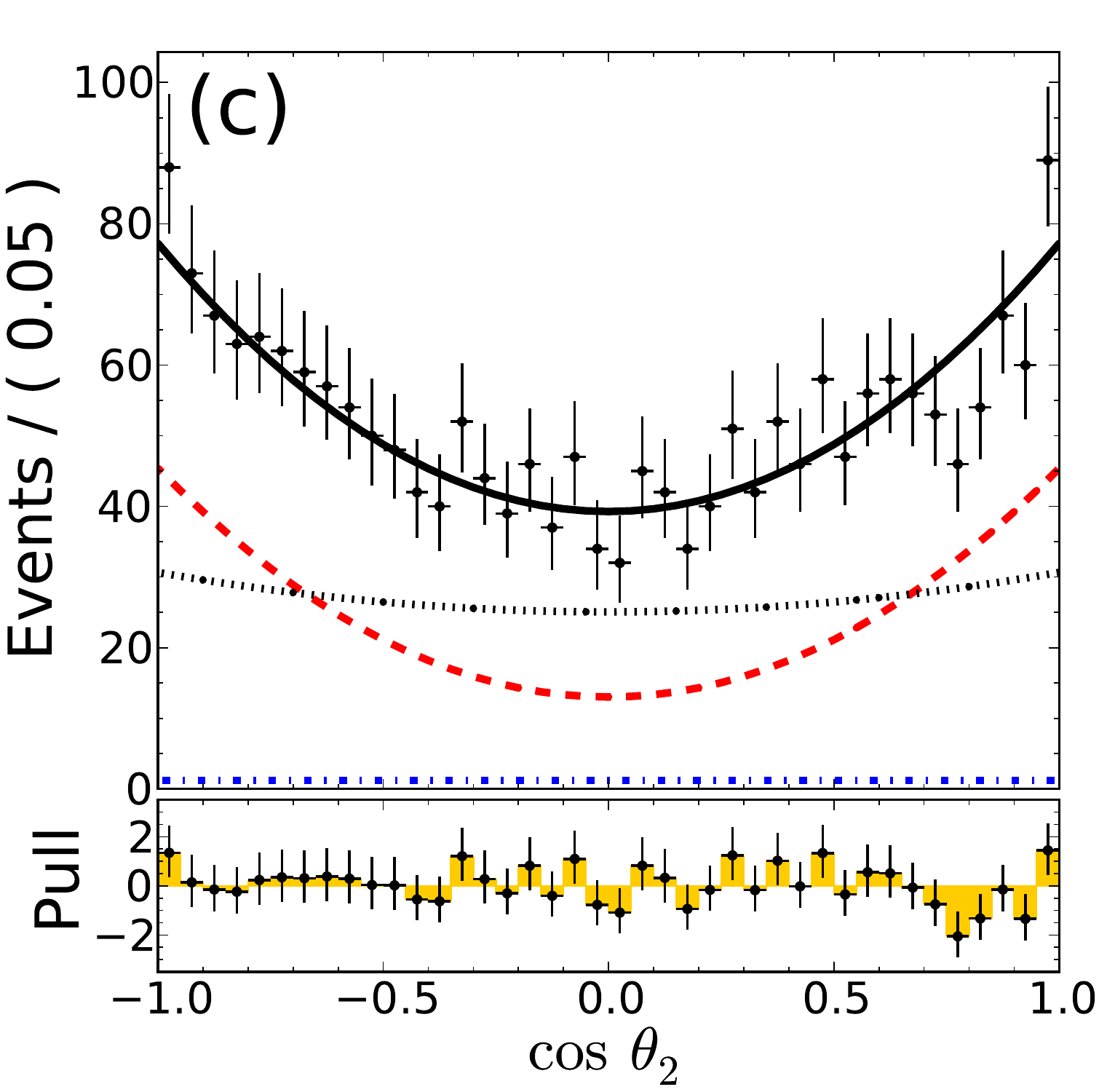}
	\includegraphics[width=0.245\linewidth]{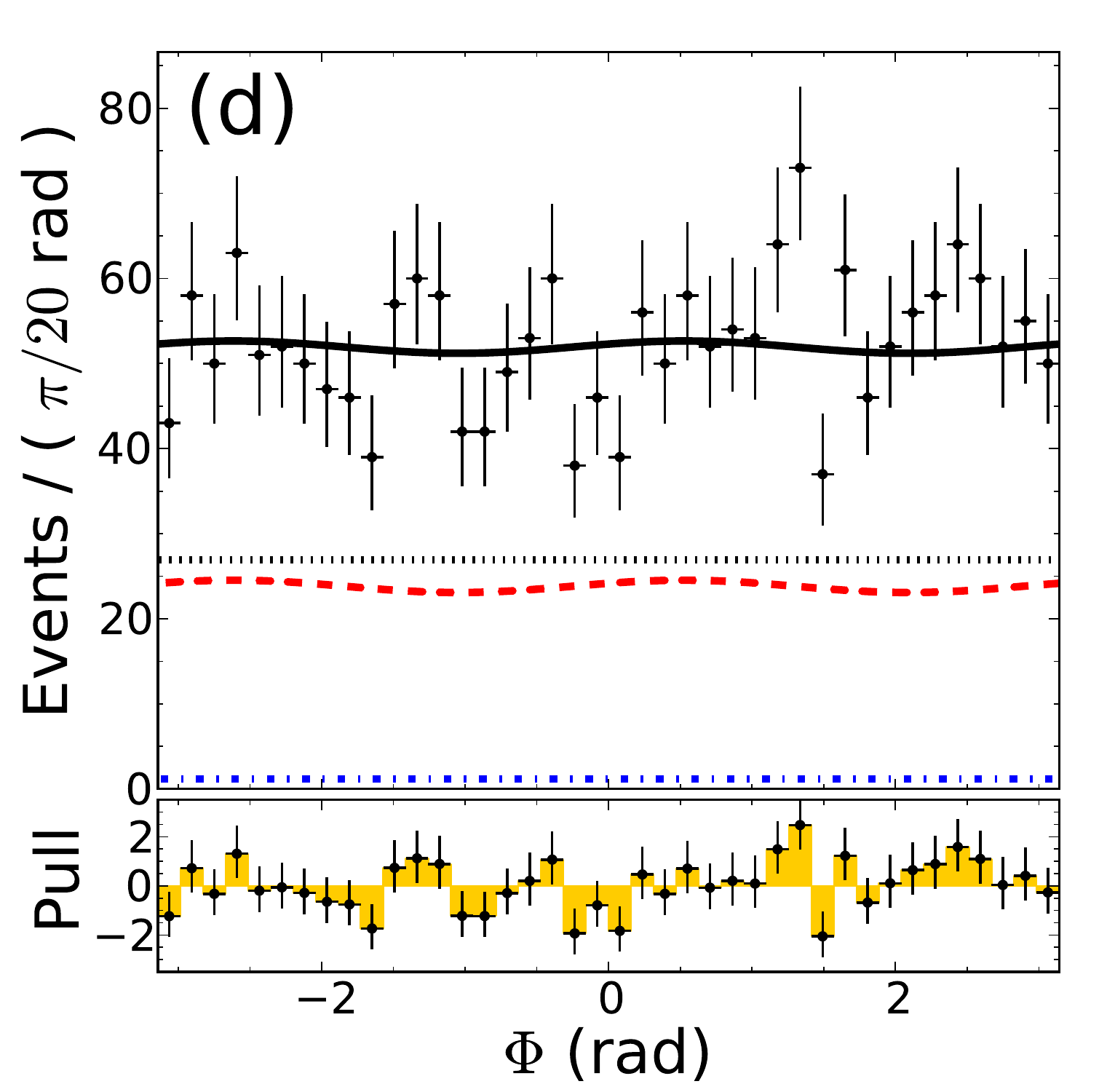}
	\caption{Projections onto the observables (a) $\mkpi$, (b) $\helthetaone$, (c) $\helthetatwo$, and (d) $\helphi$ for $\Bdecay$ and $\BdecayCC$ combined. The data distributions are shown by black markers with error bars whereas the overall fit function, combinatorial background, signal and peaking background are shown with solid black, dotted black, dashed red and dash-dotted blue curves, respectively. For each projection, the data points and fit projections are shown after a signal-enhancing selection (see text) on $\mbc$, $\deltae$, $\mkk$ and $\cnb$ shown in Fig.~\ref{fig:DATA_projections_signal_range}.}
	\label{fig:DATA_projections_signal_range_massangular}
\end{figure*}

We further consider the possibility of interference effects between the S- and P-wave $K^+K^-$ components in $\Bdecayf$ and $\BdecayP$ decays. In principle, these interference effects can be treated in a similar manner to those in the invariant $K^+\pi^-$ mass by including all amplitudes with their corresponding angular distributions in the matrix element and leading to a full partial wave analysis of $B^0 \to (K^+K^-) (K^+ \pi^-)^*$ decays. To estimate the systematic uncertainty from neglecting this interference, we include the interference term of $\Bdecayf$ and $\BdecayP$ decays into the fit model. We neglect the interference of $\Bdecayf$ with $\BdecayS$ and $\BdecayD$, as there is little overlap between these channels. We use the difference of this fit with respect to the nominal fit as the systematic uncertainty due to interference effects.

A charge bias in the reconstruction efficiency that would affect the relative yields between $B^0$ and $\bar{B}^0$ is estimated following the procedures in the analysis of $D^+ \rightarrow K_S^0 K^+$~\cite{BN1211} and $D^+ \rightarrow K_S^0 \pi^+$~\cite{BN1276} decays; it is found to be consistent with zero. We assign the uncertainty of $1.2\%$ in this estimate as a systematic uncertainty.

All systematics that neither cancel nor are negligible are evaluated in terms of their effects on the triple-product correlations and the parameters defined in Tables~\ref{tab:physics_parameter_definition} as well as on the fit fraction per partial wave, which is defined in Sec.~\ref{sec:results}. The systematic errors are summarized in Tables~\ref{tab:systematics_error_triple_product_correlations} and \ref{tab:systematics_error_physics_parameters} together with the total uncertainty by adding the individual errors in quadrature. All parameters in Table~\ref{tab:systematics_error_physics_parameters} that enter the calculation of the branching fraction are also summarized with relative errors.

\section{Results}
\label{sec:results}

We observe a signal yield of $N_{\rm sig} = 1112 \pm 40$ events, a peaking background yield of $N_{\rm peak} = 140 \pm 19$ events, and a combinatoric background yield of $N_{\rm comb} = 14522 \pm 122$ events, where the errors are statistical only. To illustrate the fit result, we show projections of the fit onto various discriminating observables in Figs.~\ref{fig:DATA_projections_signal_range} and~\ref{fig:DATA_projections_signal_range_massangular}. In each plot of Fig.~\ref{fig:DATA_projections_signal_range}, we apply a signal-enhancing requirement on the other three observables; such requirements applied for each observable are $\mbc > 5.27 \text{~GeV}$, $-40 \text{~MeV} < \deltae < 40 \text{~MeV}$, $1.01\text{~GeV} < \mkk < 1.03\text{~GeV}$ and $\cnb > -3$. In each plot of Fig.~\ref{fig:DATA_projections_signal_range_massangular}, we apply a signal-enhancing requirement on all four observables shown in Fig.~\ref{fig:DATA_projections_signal_range}.

\begin{table*}[ht]
\caption{Summary of the results on the $\BdecayK$ system. See Table~\ref{tab:physics_parameter_definition} and Eq.~(\ref{eqn:fit_fraction}) for the parameter definition. In this table, we give the fit fraction $FF_J$ per partial wave instead of the branching fraction $\mathcal{B}_J$, which is given in Table~\ref{tab:FINAL_branching_fractions} together with the yields per partial wave. The first error is statistical and the second due to systematics.}
\label{tab:FINAL_fit_results_physics_parameters}
\begin{tabular}{lccc} \hline\hline
 & $\phi \Swave$ & $\phi \Pwave$ & $\phi \Dwave$ \\
Parameter & $J=0$ & $J=1$ & $J=2$ \\ \hline
       $FF_J$ & $\phantom{-}0.273 \pm 0.024 \pm 0.021$ & $\phantom{-}0.600 \pm 0.020 \pm 0.015$ & $\phantom{-}0.099 ^{+0.016}_{-0.012} \pm 0.018$\\
     $f_{LJ}$ & $\cdots$ & $\phantom{-}0.499 \pm 0.030 \pm 0.018$ & $\phantom{-}0.918 ^{+0.029}_{-0.060} \pm 0.012$\\
$f_{\perp J}$ & $\cdots$ & $\phantom{-}0.238 \pm 0.026 \pm 0.008$ & $\phantom{-}0.056 ^{+0.050}_{-0.035} \pm 0.009$\\
$\phi_{\parallel J}$ (rad) & $\cdots$ & $\phantom{-}2.23 \pm 0.10 \pm 0.02$ & $\phantom{-}3.76 \pm 2.88 \pm 1.32$\\
    $\phi_{\perp J}$ (rad) & $\cdots$ & $\phantom{-}2.37 \pm 0.10 \pm 0.04$ & $\phantom{-}4.45 ^{+0.43}_{-0.38} \pm 0.13$\\
       $\delta_{0J}$ (rad) & $\cdots$ & $\phantom{-}2.91 \pm 0.10 \pm 0.08$ & $\phantom{-}3.53 \pm 0.11 \pm 0.19$\\
      $\mathcal{A}_{CPJ}$ & $\phantom{-}0.093 \pm 0.094 \pm 0.017$ & $-0.007 \pm 0.048 \pm 0.021$ & $-0.155 ^{+0.152}_{-0.133} \pm 0.033$\\
    $\mathcal{A}_{CPJ}^0$ & $\cdots$ & $-0.030 \pm 0.061 \pm 0.007$ & $-0.016 ^{+0.066}_{-0.051} \pm 0.008$\\
$\mathcal{A}_{CPJ}^\perp$ & $\cdots$ & $-0.14 \pm 0.11 \pm 0.01$ & $-0.01 ^{+0.85}_{-0.67} \pm 0.09$\\
$\Delta\phi_{\parallel J}$ (rad) & $\cdots$ & $-0.02 \pm 0.10 \pm 0.01$ & $-0.02 \pm 1.08 \pm 1.01$\\
    $\Delta\phi_{\perp J}$ (rad) & $\cdots$ & $\phantom{-}0.05 \pm 0.10 \pm 0.02$ & $-0.19 \pm 0.42 \pm 0.11$\\
       $\Delta\delta_{0J}$ (rad) & $\cdots$ & $\phantom{-}0.08 \pm 0.10 \pm 0.01$ & $\phantom{-}0.06 \pm 0.11 \pm 0.02$\\
\hline\hline 
 \end{tabular}
\end{table*}

To obtain the branching fraction per partial wave, we calculate the fit fraction $FF_J$ per partial wave $\mathcal{A}_J$, which is defined as
\begin{equation}
\label{eqn:fit_fraction}
FF_J = \frac{\int \vert \mathcal{A}_J \vert^2 }{\int \vert \mathcal{M} \vert^2} = \frac{\int \vert \mathcal{A}_J \vert^2 }{\int \vert \mathcal{A}_0 + \mathcal{A}_1 +\mathcal{A}_2\vert^2}.
\end{equation}

The fit fractions are given in Table~\ref{tab:FINAL_fit_results_physics_parameters} and their sum is $(97.2 \pm 0.7)\%$, where the error is statistical only. This indicates the presence of constructive interference between the partial waves.

From the product of signal yield and fit fraction, we obtain the yield per partial wave $N_J$, which is used to calculate the branching fraction. The results for the branching fraction are summarized in Table~\ref{tab:FINAL_branching_fractions} and the results for the polarization and $CP$ violation asymmetries are summarized in Table~\ref{tab:FINAL_fit_results_physics_parameters}. The results for $\BdecayP$ supersede our previous results; all results on $\BdecayS$, $\BdecayP$, and $\BdecayD$ are consistent with BaBar measurements~\cite{BaBar_phiK}, with smaller errors for $\BdecayS$ and $\BdecayP$.

\begin{figure}[ht]
\centering
	\includegraphics[width=0.49\linewidth]{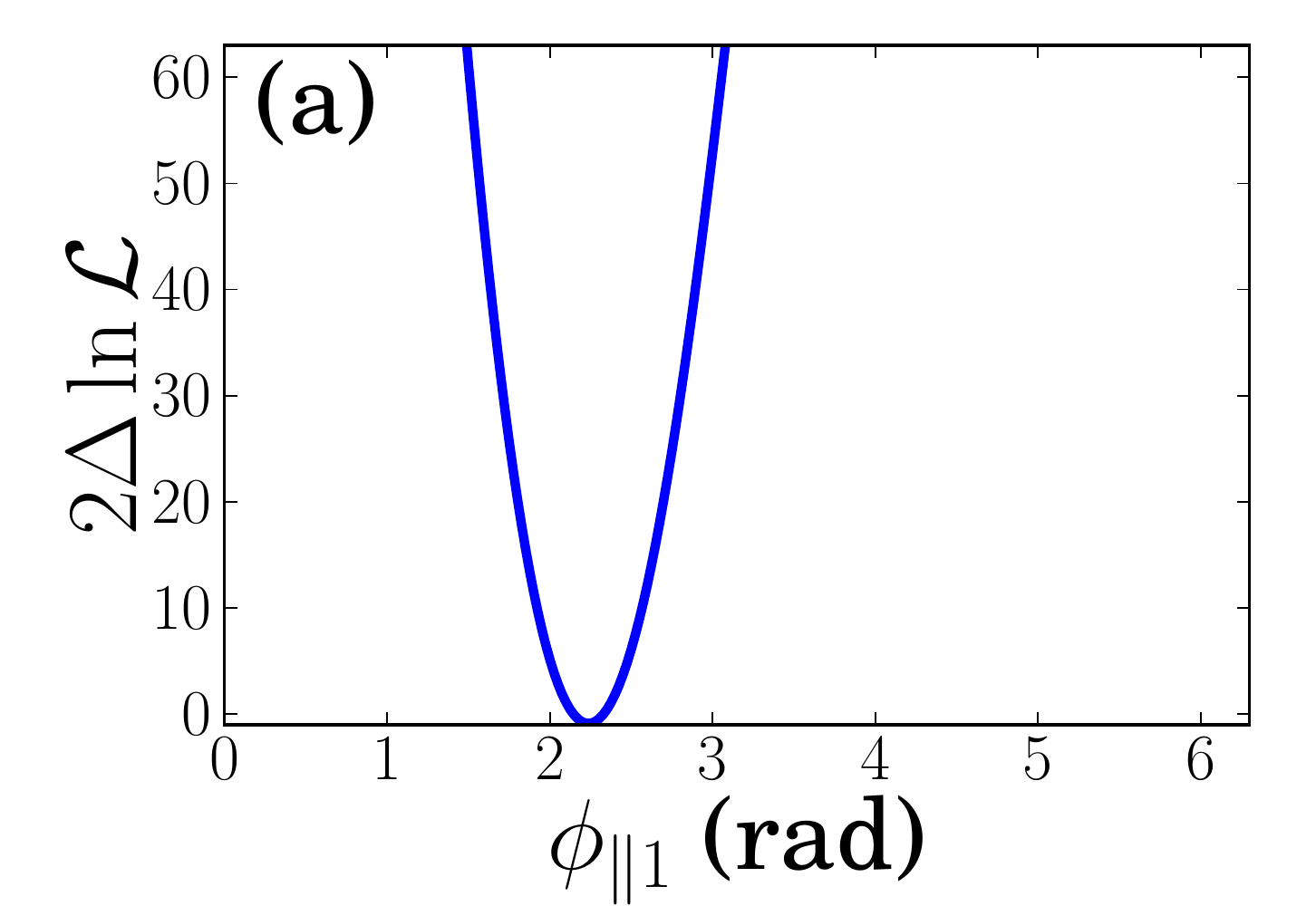}
	\includegraphics[width=0.49\linewidth]{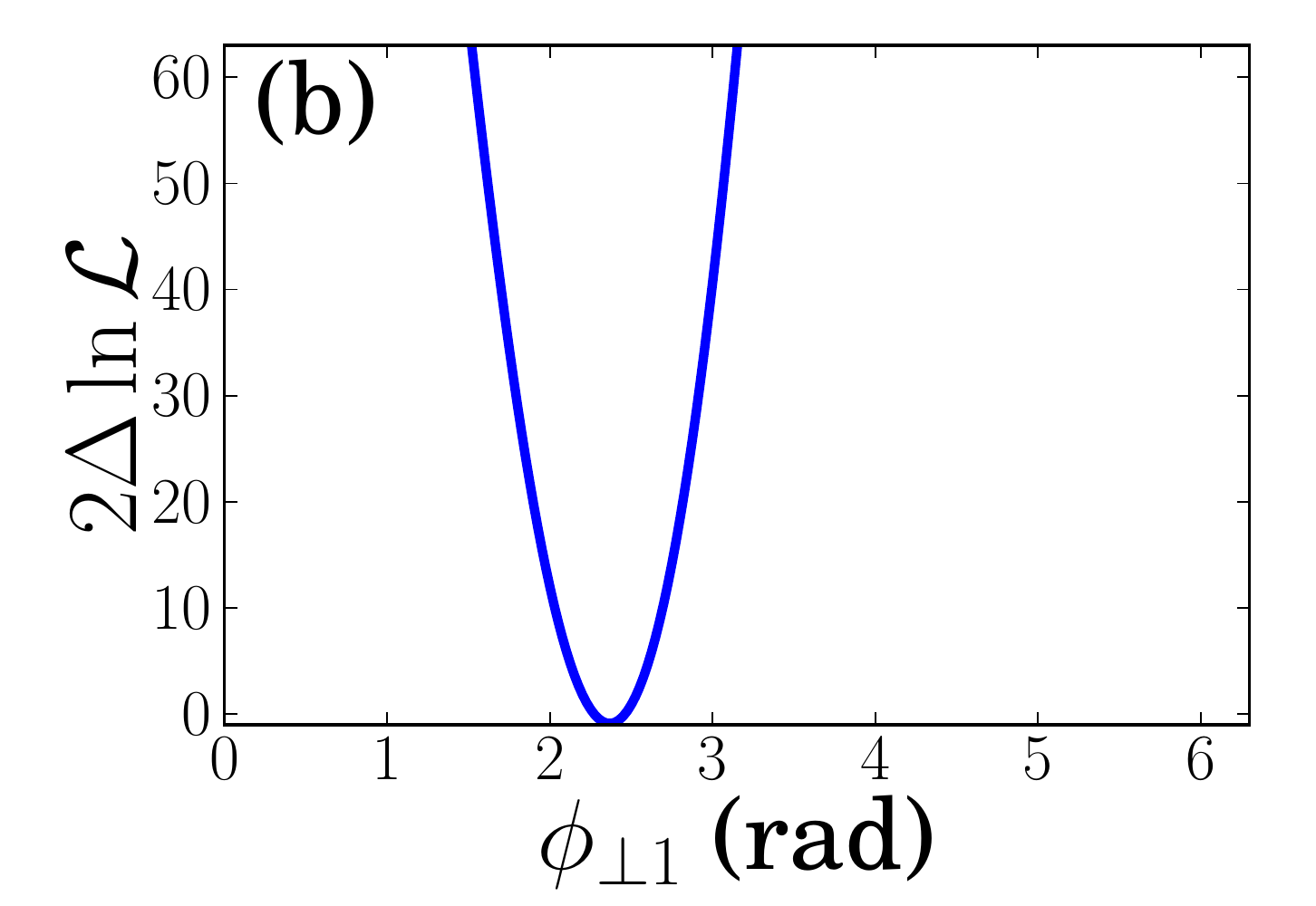}
	\caption{Scan of the negative log likelihood as function of (a) $\phi_{\parallel 1}$ and (b) $\phi_{\perp 1}$. One single discrete solution is found for each of the two phases.}
	\label{fig:DATA_likelihoodscan_phase_parallel_perp}
\end{figure}

We resolve the ambiguity in the phase parameters $\phi_{\parallel 1}$ and $\phi_{\perp 1}$ from our previous measurement. In Fig.~\ref{fig:DATA_likelihoodscan_phase_parallel_perp}, we show a scan of the negative log-likelihood as a function of $\phi_{\parallel 1}$ and $\phi_{\perp 1}$, each of which shows a single solution. We also confirm the large longitudinal polarization fraction in the decay $\BdecayD$ observed by BaBar~\cite{BaBar_phiK}. In general, all parameters related to $CP$ violation in the S-, P-, and D-wave components are consistent with its absence.

\begin{table}[ht]
\caption{Summary of the branching fraction results for the $\BdecayK$ system. The result for $\BdecayS$ is quoted for $\mkpi < 1.55~{\rm GeV}$. The first error is statistical and the second due to all systematics. The error on $\epsilon_{\text{reco},J}$ is due to MC statistics only. For the overall efficiency $\epsilon_J$, defined as $\epsilon_{\text{reco},J}$ times daughter branching fractions, the error is due to MC statistics and daughter branching fractions.}
\label{tab:FINAL_branching_fractions}
\begin{tabular}{lccc} \hline\hline
 & $\phi \Swave$ & $\phi \Pwave$ & $\phi \Dwave$ \\
Parameter & $J=0$ & $J=1$ & $J=2$ \\ \hline
$N_J$ (events) & $303 \pm 29 \pm 25 $ & $668 \pm 34 \pm 24$ & $110 ^{+18}_{-14} \pm 20$\\
$\epsilon_{\text{reco},J}$ (\%) & $28.7 \pm 0.1$ & $26.0 \pm 0.1$ & $16.3 \pm 0.1$ \\
$\epsilon_J$ (\%) & $9.4 \pm 0.1 $ & $8.5 \pm 0.1$ & $2.6 \pm 0.1$ \\
$\mathcal{B}_J$ $(10^{-6})$ & $4.3 \pm 0.4 \pm 0.4$ & $10.4 \pm 0.5 \pm 0.6$ & $5.5 ^{+0.9}_{-0.7} \pm 1.0$\\
\hline\hline 
\end{tabular}
\end{table}

Due to our requirement on $\helthetaone$ and the large longitudinal polarization in $\BdecayD$ we observe a proportionally large drop in the efficiency with respect to the other channels, which results in larger statistical uncertainties on the related parameters.

The results on the triple-product correlations in $\BdecayP$ are summarized for $B^0$ and $\bar{B}^0$, together with the asymmetries, in Table~\ref{tab:FINAL_triple_product_correlations}. They are consistent with SM predictions of no $CP$ violation.

\begin{table}[ht]
\caption{Triple-product correlations obtained from the weights of the $\BdecayP$ partial wave. The first error is statistical and the second due to systematics.}
\label{tab:FINAL_triple_product_correlations}
\begin{tabular}{lcc} \hline\hline
             & $A_{T}^{0}$ & $A_{T}^{\parallel}$ \\ \hline
$B^0$        & $\phantom{-}0.273 \pm 0.039 \pm 0.010$ & $\phantom{-}0.015 \pm 0.029 \pm 0.006$ \\
$\bar{B}^0$  & $\phantom{-}0.210 \pm 0.039 \pm 0.014$ & $\phantom{-}0.050 \pm 0.029 \pm 0.011$ \\
$\mathcal{A}_{T}^{0/\parallel}$ & $\phantom{-}0.13 \pm 0.12 \pm 0.02$ & $-0.55 \pm 0.60 \pm 0.52$\\
\hline\hline 
 \end{tabular}
\end{table}

\section{Conclusion}
In summary, we have performed a partial wave analysis of the $\BdecayK$ system and measured branching fraction and polarization parameters for the S-, P-,\linebreak and D-wave contribution from $\BdecayS$, $\BdecayP$, and $\BdecayD$, respectively. We have resolved all phase ambiguities present in our previous polarization measurements in these decays. We have further searched for $CP$ violation in these decays. Results are summarized in Tables~\ref{tab:FINAL_fit_results_physics_parameters}, \ref{tab:FINAL_branching_fractions} and \ref{tab:FINAL_triple_product_correlations}. All parameters related to $CP$ violation are consistent with its absence in the studied decays and no evidence for new physics is found.

\section*{Acknowledgments}

We thank the KEKB group for the excellent operation of the
accelerator; the KEK cryogenics group for the efficient
operation of the solenoid; and the KEK computer group,
the National Institute of Informatics, and the 
PNNL/EMSL computing group for valuable computing
and SINET4 network support.  We acknowledge support from
the Ministry of Education, Culture, Sports, Science, and
Technology (MEXT) of Japan, the Japan Society for the 
Promotion of Science (JSPS), and the Tau-Lepton Physics 
Research Center of Nagoya University; 
the Australian Research Council and the Australian 
Department of Industry, Innovation, Science and Research;
Austrian Science Fund under Grant No. P 22742-N16;
the National Natural Science Foundation of China under
contract No.~10575109, 10775142, 10875115 and 10825524; 
the Ministry of Education, Youth and Sports of the Czech 
Republic under contract No.~MSM0021620859;
the Carl Zeiss Foundation, the Deutsche Forschungsgemeinschaft
and the VolkswagenStiftung;
the Department of Science and Technology of India; 
the Istituto Nazionale di Fisica Nucleare of Italy; 
The BK21 and WCU program of the Ministry Education Science and
Technology, National Research Foundation of Korea Grant No.\ 
2010-0021174, 2011-0029457, 2012-0008143, 2012R1A1A2008330,
BRL program under NRF Grant No. KRF-2011-0020333,
and GSDC of the Korea Institute of Science and Technology Information;
the Polish Ministry of Science and Higher Education and 
the National Science Center;
the Ministry of Education and Science of the Russian
Federation and the Russian Federal Agency for Atomic Energy;
the Slovenian Research Agency;
the Basque Foundation for Science (IKERBASQUE) and the UPV/EHU under 
program UFI 11/55;
the Swiss National Science Foundation; the National Science Council
and the Ministry of Education of Taiwan; and the U.S.\
Department of Energy and the National Science Foundation.
This work is supported by a Grant-in-Aid from MEXT for 
Science Research in a Priority Area (``New Development of 
Flavor Physics''), and from JSPS for Creative Scientific 
Research (``Evolution of Tau-lepton Physics'').


\begin{thebibliography}{99}

\bibitem{C}
N.~Cabibbo, Phys. Rev. Lett. {\bf 10}, 531 (1963).

\bibitem{KM}
M.~Kobayashi and T.~Maskawa, Prog. Theor. Phys. {\bf 49}, 652 (1973).

\bibitem{HFAG}
Y.~Amhis {\it et al.} (Heavy Flavor Averaging Group), 
arXiv:1207.1158 [hep-ex] and online update at 
http://www.slac.stanford.edu/xorg/hfag.

\bibitem{Belle_phiK}
K.-F.~Chen {\it et al.} (Belle Collaboration), Phys. Rev. Lett. {\bf 94}, 221804 (2005).

\bibitem{BaBar_phiK}
B.~Aubert {\it et al.} (BaBar Collaboration), Phys. Rev. D {\bf 78}, 092008 (2008).

\bibitem{Polarization}
A.~Datta {\it et al.}, Phys. Rev. D {\bf 77}, 114025 (2008) and references therein for discussions of SM and new-physics explanations of $f_T/f_L$.

\bibitem{CC}
Throughout this paper, the inclusion of the charge-conjugate mode decay is implied unless otherwise stated.

\bibitem{alpha} Another naming convention $\beta$ ($= \phi_1$) is also used in the literature.

\bibitem{LASS}
D.~Aston {\it et al.} (LASS Collaboration), Nuclear Phys. B {\bf 296}, 493 (1988).

\bibitem{BaBar_highmass_states}
B.~Aubert {\it et al.} (BaBar Collaboration), Phys. Rev. D {\bf 76}, 051103(R) (2007).

\bibitem{PDG}
J.~Beringer {\it et al.} (Particle Data Group), Phys. Rev. D {\bf 86}, 010001 (2012).

\bibitem{Wigner}
E.~P.~Wigner, Phys. Rev. {\bf 98}, 145 (1955).

\bibitem{TripleProduct}
A.~Datta and D.~London, Int. J. Mod. Phys. A {\bf 19}, 2505 (2004).

\bibitem{TripleProduct_note}
We take the definitions of $A_T^0 = A_T^{(1)}$ and $A_T^\parallel = A_T^{(2)}$, but $\bar{A}_T^0 = -\bar{A}_T^{(1)}$ and $\bar{A}_T^\parallel = -\bar{A}_T^{(2)}$. The variables $A_T^{(1)}$ and $A_T^{(2)}$ are defined in Ref.~\cite{TripleProduct}.

\bibitem{KEKB}
S.~Kurokawa and E.~Kikutani, Nucl. Instrum. Methods Phys. Res. Sect. A {\bf 499}, 1 (2003), and other papers included in this Volume;   T.~Abe {\it et al.}, Prog. Theor. Exp. Phys. (2013) 03A001 and following articles up to 03A011.
  
\bibitem{Belle}
A.~Abashian {\it et al.} (Belle Collaboration), Nucl. Instrum. Methods Phys. Res. Sect. A {\bf 479}, 117 (2002); also see detector section in J.~Brodzicka {\it et al.}, Prog. Theor. Exp. Phys. (2012) 04D001.

\bibitem{svd2} Z.~Natkaniec {\it et al.} (Belle SVD2 Group), Nucl. Instrum. Methods Phys. Res. Sect. A {\bf 560}, 1 (2006).

\bibitem{NeuroBayes}
M.~Feindt and U.~Kerzel, Nucl. Instrum. Methods Phys. Res. Sect. A {\bf 559}, 190 (2006).

\bibitem{SFW}
The Fox-Wolfram moments were introduced in
G.~C.~Fox and S.~Wolfram, Phys. Rev. Lett. {\bf 41}, 1581 (1978).
The Fisher discriminant used by Belle, based on modified Fox-Wolfram
moments (SFW), is described in 
K.~Abe {\it et al.} (Belle Collaboration), Phys. Rev. Lett. {\bf 87},
101801 (2001) and
K.~Abe {\it et al.} (Belle Collaboration), Phys. Lett. B {\bf 511}, 151
(2001). 
 
\bibitem{EvtGen}
D.~J.~Lange, Nucl. Instrum. Methods Phys. Res. Sect. A {\bf 462}, 152 (2001).

\bibitem{GEANT3}
R.~Brun {\it et al.}, GEANT 3.21, CERN Report No. DD/EE/84-1 (1987).

\bibitem{PHOTOS}
E.~Barberio and Z.~Was, Comput. Phys. Commun. {\bf 79}, 291 (1994).

\bibitem{CAT}
M.~Feindt and M.~Prim, Nucl. Instrum. Methods Phys. Res. Sect. A {\bf 698}, 84 (2013).

\bibitem{BES}
M.~Ablikim {\it et al.} (BES Collaboration), Phys. Lett. B {\bf 607}, 243 (2005).

\bibitem{F76}
S.~M.~Flatte, Phys. Lett. B {\bf 63}, 224 (1976).

\bibitem{Argus}
H.~Albrecht {\it et al.} (ARGUS Collaboration), Phys. Lett. B {\bf 241}, 278 (1990).

\bibitem{Minuit} F.~James {\it et al.}, MINUIT, Function Minimization and Error Analysis CERN Program Library.

\bibitem{RooFit} W.~Verkerke and D.~Kirkby, The RooFit Toolkit for Data Modeling, Proceedings from CHEP03 (2003)

\bibitem{ROOT} R.~Brun and F.~Rademakers, Nucl. Inst. Meth. A {\bf 389}, 81 (1997).

\bibitem{KKinterference} Y.~Nakahama {\it et al.} (Belle Collaboration), Phys. Rev. D {\bf 82}, 073011 (2010).

\bibitem{BN1211} 
B.~R.~Ko {\it et al.} (Belle Collaboration), Phys. Rev. Lett. {\bf 109}, 021601 (2012) and Phys. Rev. Lett. {\bf 109}, 119903(E) (2012).

\bibitem{BN1276} 
B.~R.~Ko {\it et al.} (Belle Collaboration), JHEP {\bf 02}, 98 (2013).

\end{thebibliography}
\end{document}